\documentstyle[12pt]{article}

\newcommand{\be}{\begin{eqnarray}}
\newcommand{\ee}{\end{eqnarray}}
\newcommand{\ben}{\begin{eqnarray*}}
\newcommand{\enn}{\end{eqnarray*}}
\newcommand{\gb}{\bar{g}}

\newcommand{\f}{\frac}
\newcommand{\al}{\alpha}
\newcommand{\rc}{\gamma^{c}}
\newcommand{\ab}{\bar{\alpha}}
\newcommand{\az}{\alpha_{0}}

\newcommand{\bee}{\begin{eqnarray}}
\newcommand{\en}{\end{eqnarray}}
\newcommand{\ka}{\kappa}
\newcommand{\ga}{\gamma}
\newcommand{\ba}{\beta}

\newcommand{\ra}{\rightarrow}
\newcommand{\ro}{\gamma_{00}}

\begin{document}

\renewcommand{\thepage}{\roman{page}}

\thispagestyle{empty}

\textheight 200mm\textwidth 150 mm
\hoffset=-0.5cm
\voffset=-1.7cm
\renewcommand{\baselinestretch}{1.2}

\begin{titlepage}   
\begin{flushright}
EFI 96-16 \\
hep-th/9607045 

\end{flushright}

\begin{center}
\vskip 0.3truein

{\large{\bf{ASYMPTOTIC LIMITS AND SUM RULES}}}
\vskip 0.15truein

{\large{\bf{FOR PROPAGATORS IN QUANTUM }}}
\vskip 0.15truein

{\large{\bf{CHROMODYNAMICS}}}
\footnote{Work supported in part by the National
Science Foundation, Grant PHY 91-23780. \\
E-mail: jackxu@yukawa.uchicago.edu}
\vskip 0.6truein

{\large{ Wentao Jack Xu}}
\vskip 0.2truein

Enrico Fermi Institute and Department of Physics

University of Chicago

Chicago, Illinois 60637, USA
\end{center}
\vskip 0.5truein
\end{titlepage}

\newpage\setcounter{page}{2}
\tableofcontents
\newpage


\section*{Abstract}

In gauge field theories with asymptotic freedom, the short distance properties of Green's functions can be 
obtained on the basis of weak coupling perturbation expansions.
Within this framework, the large momentum behavior of the structure functions for gluon, quark and ghost propagators is derived.
The limits are found for general, covariant, linear gauges, and in all directions of the complex $k^2-$plane. Except for the coefficients, the functional forms of the leading asymptotic terms for the various structure functions are independent of the gauge parameter.
They are determined exactly in terms of one-loop expressions (two-loop expressions in cases where one-loop terms vanish).
With the exception of the Landau gauge, the asymptotic expressions for the gauge field propagator play an important r\^{o}le for the corresponding limits of quark and ghost propagators.
For {\it all} gauges considered, it is the sign of the one-loop anomalous dimension coefficient of the gluon field in Landau gauge (as a fixed point of the gauge parameter) which is of considerable relevance for the asymptotics of the various propagators.

The bounds obtained from the asymptotic expressions, together with the analytic properties of the structure functions, generally lead to un-subtracted dispersion representations. In special cases, for a limited number of flavors, sum rules are obtained for the discontinuities along the real axis. The sum rule for the gluon propagator is a generalization of the superconvergence relation derived previously in the Landau gauge.

\newpage
\renewcommand{\thepage}{\arabic{page}}\setcounter{page}{1}

\section{Introduction}

For field theories with asymptotic freedom, it is possible to describe the Green's functions at short  distances in terms of weak coupling perturbation expansions. Consequently, the theory is well defined in this limit \cite{gw}. On the other hand, the long distance behavior involves elements which are not seen in the asymptotic expansions. These elements determine the phase structure of the theory, like the possible confinement of gluons and quarks in quantum chromodynamics (QCD) for limited number of quark flavors. Since a direct calculational approach does not appear to be feasible in the infrared region, it is of interest to use indirect methods in order to obtain some information.
For propagators and vertex functions, one can obtain interesting results using analytic properties resulting from locality and spectral conditions, together with asymptotic expansions for large momenta, which can be derived with the help of renormalization group methods.

It is the purpose of this paper to present results for propagators of non-Abelian gauge theories, like QCD, in general, covariant, linear gauges. We consider the structure functions of gluon, quark and ghost propagators, using the massless theory with an Euclidean renormalization mass as the only dimensionful parameter. Although there are no intrinsic masses in the theory, the dynamical generation of mass gaps is not excluded.

For the Landau gauge ($\al=0$), the asymptotic properties of the gauge field propagator have been 
obtained in \cite{oz, ooz, 128}. A detailed discussion of the distribution aspects of 
the structure functions may be found in \cite{oz}. A brief description of the asymptotic 
expansions for gluon and quark propagators has been given in the letters \cite{ox} 
and \cite{oxq} respectively.

Our results are based on general principles only. We use Lorentz covariance and minimal spectral conditions as formulated in the state space of indefinite metric of the gauge theory considered \cite{ko}. This is sufficient in order to have the structure functions of propagators as boundary values of analytic functions, which are regular in the complex $k^2-$plane, with cuts along the positive real axis only \cite{kl}. We then use renormalization group methods \cite{OZ7}, which make it possible to derive asymptotic expressions for the structure functions in all directions of the complex $k^2-$plane, including those parallel to the positive real axis. For some results, we make use of the assumption that the exact Green's functions are connected with the formal perturbation expansion in the limit of vanishing coupling, at least as far as the first few terms are concerned.

Except for coefficients, we find that the leading terms in the asymptotic expansions are independent of the gauge parameter, and determined by one-loop information (or two-loop in cases where one-loop terms vanish). The one-loop coefficient $\beta_0$ in the weak coupling expansion of the renormalization group function $\beta$ and the corresponding one-loop anomalous dimension coefficients for gluon, quark and ghost fields play an important r\^{o}le for the asymptotic expressions. In all cases, there is sufficient boundedness for the existence of unsubtracted dispersion representations of the structure functions. Even dipole representations are generally possible. In special interesting cases, we find that the discontinuities of the structure functions obey sum rules which are generalization of the superconvergence relation obtained in \cite{oz, 128} for the gauge field propagator in the Landau gauge. In most cases, and for {\it general gauges}, the existence of these sum rul!
es
depends importantly upon the sign of the one-loop anomalous dimension coefficient $\gamma_{00}$
of the gauge field in the Landau gauge. In recent generalizations of these sum rules to $N = 1 $
SUSY theories with asymptotic freedom \cite{on,of}, it was shown that this coefficient $\ga_{00}$ is directly related to the one-loop $\beta-$ function coefficient of the dual theory \cite{ws,s}. In particular in the Landau gauge, the sum rules play an important r\^{o}le for the problem of confinement \cite{o,np}, and the connection mentioned above for $N=1$ SUSY models allows comparison with results for the phase structure of the theory \cite{on}, which can be obtained on the basis of duality \cite{s}. For the gauge field propagator, the dipole representation mentioned above, together with the detailed asymptotic limit of the discontinuity, is important for the presence of an approximately linear potential between static quarks and anti-quarks \cite{126}. This potential is indicated provided the number of flavors is sufficiently small so that coefficient $\ga_{00}$ is negative.

In this paper, we concentrate on the derivation of asymptotic expressions for the various structure functions and of related sum rules. Possible applications will be considered elsewhere. In particular, generalizations to SUSY theories and their dual maps are of interest.

We define the gauge parameter $\al$ by writing the gauge fixing term of the theory in the form \cite{ko}
\be
{\cal L}_{GF} = -B\cdot (\partial_\mu A^\mu) +
\frac{\alpha}{2} B\cdot B,
\label{GF}
\en
where the auxiliary, hermitian $B-$field satisfies the equations
\be
\partial_\mu A^\mu - \al B =0,\\
\Box B =0.
\en
With Eq.(\ref{GF}), the gauge parameter $\al$ is defined so that $\al=0$ corresponds to the Landau gauge and $\al=1$ to the Feynman gauge.

In the following, we will often use the language of QCD. The structure 
function $D(k^2+i0)$ of the gluon propagator is defined by
\begin{eqnarray}
D_F^{\mu\nu\rho\sigma}(k)&=&\int dx e^{ikx}   \langle 0 | T A^{\mu \nu}_{a}(x)
A^{\varrho\sigma}_{b}(0) |0\rangle \cr
-iD_F^{\mu\nu\rho\sigma}(k)&=&\delta_{ab} D (k^2 + i0)
\left( k^\mu k^\varrho g^{\nu \sigma} -
k^\mu k^\sigma g^{\nu \varrho}
+ k^\nu k^\sigma g^{\mu\varrho} - k^\nu k^\varrho g^{\mu\sigma}\right)
\label{defD}
\end{eqnarray}
with $A^{\mu \nu}\equiv\partial^\mu A^\nu - \partial^\nu A^\mu$.
The use of $A^{\mu\nu}(x)$ eliminates the longitudinal part. 

For the quark propagator, we write 
\be
-S_{F}(k) &=& \int d^4x \exp^{ik\cdot x}
\langle 0|T\psi(x)\bar{\psi}(0)|0\rangle,\cr
-iS_F(k)&=& A(k^2+i0)\sqrt{-k^2} + B(k^2+i0)\ga\cdot k,
\label{sq}
\en
and the ghost structure function is 
\be
-D_c(k) &=& \int d^4x \exp^{ik\cdot x}
\langle 0|T\bar{\eta}(x)\eta(0)|0\rangle\cr
-iD_c(k)& =& D_c(k^2+i0).
\label{D_c}
\en

\bigskip

The limits we have obtained for various structure functions, together with the analytic properties following from Lorentz invariance, allow one immediate conclusion: these functions cannot be non-trivial and entire, but must have singularities on the positive real $k^2-$axis. This is compatible with a strict framework of tempered distributions. The singularities are generally associated with unphysical and/or confined excitations. There is no problem with the existence of corresponding asymptotic states in the general state space of the theory. Within the framework of BRST-quantization \cite{BRST, ko}, these states correspond to quartet representations of the BRST-algebra, and hence are not elements of the physical state space \cite{o}. In contrast, as has been discussed in \cite{137, 140, 133}, it turns out that physical (hadronic) amplitudes do not have any singularities which are associated with the quark-gluon structure. This includes anomalous thresholds which normally 
describe the composite structure 
with {\it confined} constituents, even loosely bound heavy quarks.

\section{Gluon Propagator}

In this section, we derive the large $k^2$ asymptotic expansion for the gauge field propagator in all directions in the complex $k^2-$plane \cite{ox}. To obtain this result, we will apply the renormalization group symmetry which relates the large momentum limit of the structure function to its weak coupling limit. The symmetry can be expressed as a singular nonlinear differential equation that describes how the structure function changes in terms of the  running coupling constant. Before deriving the asymptotic expansions in all directions in the complex $k^2-$plane, we first consider the case when $k^2\ra-\infty$ along the negative real axis. In this case, the  running coupling constant is real and positive, and the structure function is analytic and real. 
The renormalization group equation is then solved by a change of variables 
in terms of which the singular differential equation is transformed into one that satisfies the Lipschitz condition
 and is solvable by power series. The solution shows that the large $k^2$ limit of the structure function depends on the sign of the constant 
$\xi=\frac{\ga_{00}}{\beta_0}$. These results for $k^2\ra-\infty$ are  then extended to all directions in the complex $k^2-$plane. Important for this extension are the analyticity of the structure function in the cut $k^2-$ plane and the assumption that the exact Green's functions are connected to expansions in perturbation theory as the coupling constant approaches zero, at least as far as the first order term is concerned.

\subsection{Renormalization group equation}

The gauge field satisfies the following renormalization group equation
\be
A_\mu(x, g', \alpha', \ka') = \sqrt{Z_3}A_\mu(x, g, \alpha, \ka).
\label{az}
\en
Here
\be
Z_3 &=& Z_3\left(\frac{\ka'^2}{\ka^2},g,\al\right),\nonumber\\
g' &=& \gb\left(\frac{\ka'^2}{\ka^2},g\right) ,\nonumber\\
\al'&=&\bar{\al}\left(\frac{\ka'^2}{\ka^2},g,\al\right)=\al R^{-1}
\left(\frac{\ka'^2}{\ka^2},g,\al\right) , 
\label{4}
\en
where $R\left(\frac{k^2}{\ka^2},g,\al\right) = -k^2D(k^2, \ka^2, g,\al)$ is the
structure function of the 
transverse gauge field propagator with
the normalization
\be
R(1,g,\al) = 1.
\label{normlz}
\en  
For convenience, we will use the dimensionless structure function $R$ in place of $D$ when deriving large $k^2$ expansion. 

In Eq.(\ref{az}),  the factor $Z_3$ describes how the field operator transforms when the coupling constant $g^2$ and the gauge parameter $\al$ change according to the function $\gb$ and $\ab$. These functions are also called the running coupling constant and the running gauge parameter respectively.  Eq.(\ref{az}), together with the definition of the gauge structure function, Eq.(\ref{defD}), implies that
\be
R\left(\frac{k^2}{\ka^2},g,\al\right) = Z_3^{-1}\left(\frac{\ka'^2}{\ka^2}, g, \al\right)R\left(\frac{k^2}{\ka'^2},g',\al'\right) .
\label{Z}
\en
Setting $k^2=\ka'^2$, the normalization condition Eq.(\ref{normlz}) then 
becomes
\be 
Z_3^{-1}\left(\frac{\ka'^2}{\ka^2}, g, \al\right) = R\left(\frac{\ka'^2}{\ka^2},g,\al\right) .
\en
Substituting this result back to Eq.(\ref{Z}), we get 
the renormalization group equation in terms of the dimensionless structure function:
\be
R\left(\frac{k^2}{\ka^2},g,\al\right) = R\left(\frac{\ka'^2}
{\ka^2},g,\al\right)R\left(\frac{k^2}{\ka'^2},g',\al'\right) .
\label{7}
\en

This equation  describes the renormalization group symmetry of the dimensionless structure function. It will be used later to extend the expansion along the negative real $k^2-$ axis to all directions in the complex $k^2-$plane. For now, we would like to obtain a differential renormalization group equation in a form which is convenient for our purpose.

By differentiating Eq.(\ref{7}) with respect to $k^2$ and setting $\ka^2 = \ka'^2$, we obtain the differential equation
\be
u\frac{\partial R(u,g,\al)}{\partial u} = \ga(\gb^2,\bar\al)R(u,g,\al) ,
\label{RGE}
\en
where $u \equiv k^2/\ka^2,$ and 
\be
\ga(g^2,\al)\equiv u\frac{\partial R(u,g,\al)}{\partial u}|_{u=1}
\en
is the anomalous dimension of the gluon field, and the functions $\gb$ and $\ab$ are given  in Eq.(\ref{4}). In general, the exact form of the $\gamma$-function is unknown, but its asymptotic expansion for $g\ra0$ and {\it fixed} $\al$ can be obtained from the perturbation theory, and it is given by
\be
\gamma (g^2,\alpha) \simeq \gamma_0 (\alpha) g^2 + \gamma_1 (\alpha) g^4
+\cdots,
\label{gamma}
\en 
where
\be
\gamma_0 (\alpha) &=& \gamma_{0 0} + \alpha \gamma_{01} ,\\
\gamma_1 (\alpha) &=& \gamma_{10} +
\alpha \gamma_{11} + \alpha^2\gamma_{12}.
\en
In QCD, the perturbation theory yields  
\be
\gamma_{0 0} &=& - (16
\pi^{2})^{-1} (\frac{13}{2} -\frac{ 2}{3} N_F), \\
 \gamma_{01} &=& (16 \pi^{2})^{-1}\frac {3}{2},
\en
 where $N_F$ = number of flavors.

At first, we will consider the limit $k^2\ra-\infty$ along the negative real $k^2-$axis, where the function $R(\frac{k^2}{\ka^2},g,\al)$ is analytic and real. For this purpose, it
is convenient 
to replace the variable $u$ in Eq.(\ref{RGE}) by the function $\gb^2(u,g)$, which is defined by the initial value problem
\be
u\frac{\partial \gb^2}{\partial u} = \beta(\gb^2),\cr
\gb^2 (u=1) = g^2.
\label{dgdu}
\en
Here $\beta(g^2)$ is the renormalization group function and has the asymptotic expansion for $g^2\ra0$:
\be
\beta(g^2) \simeq \beta_0g^4 + \beta_1g^6 +\cdots.
\label{beta}
\en
In QCD, $\beta_0 = - (16 \pi^{2})^{-1} (11 - \frac{2}{3}N_F)$. The exact form of the $\beta$-function is unknown for theories without supersymmetry.

In terms of $\gb^2$, we write
\be
R(\gb^2;g^2,\al) \equiv R\left(\frac{k^2}{\ka^2},g,\al\right) ,
\label{12}
\en
and transform Eq.(\ref{RGE}) to the form 
\be
\frac{\partial R(\gb^2;g^2,\al)}{\partial\gb^2} = \frac{\ga(\gb^2,\bar\al)}{\ba(\gb^2)}R(\gb^2;g^2,\al) .
\label{16}
\en
In the case $\al\neq0$, 
since $\ab = \al R^{-1}$, we may eliminate one more dependent variable by replacing $R$ by $\ab$ and obtain the differential equation
\be
\frac{\partial \ab(\gb^2;g^2,\al)}{\partial\gb^2} = \frac{-\ab\ga(\gb^2,\bar\al)}{\ba(\gb^2)}.
\label{abar}
\en
If $\al=0$, then $\ab\equiv0$, and Eq.(\ref{16}) becomes
\be
\frac{\partial R(\gb^2;g^2,0)}{\partial\gb^2} = \frac{\ga(\gb^2,0)}{\ba(\gb^2)}R(\gb^2;g^2,0) .
\label{17}
\en
In this case, the solution can be written in the closed form: 
\be
R(\gb^2;g^2,0) = \exp\int_{g^2}^{\gb^2}dx\frac{\gamma(x,0)}{\beta(x)}.
\label{zeroa}
\ee
The $\gb^2\ra0$ asymptotics of the solution is 
given by 
\be
    R\simeq C_R(\gb^2)^\xi + \cdots,
\label{R}
\ee
 where $\xi\equiv\frac{\ga_{00}}{\beta_0}$ and
\be
C_R(g^2) &=& (g^2)^{-\xi}\exp\int_{g^2}^0dx\left(\frac{\gamma(x,0)}{\beta(x)}-\frac{\gamma_{00}}{\beta_0x}\right).
\label{CR}
\ee
Eq. (\ref{R}) and (\ref{CR}) hold for all values of $\xi\neq0$. By the asymptotic expansions in Eq.(\ref{gamma}) and (\ref{beta}), $C_R$ is finite, and furthermore $C_R>0$.

Eq.(\ref{16}) and (\ref{abar}) will be used in the following to derive the large $k^2$ expansion of the gauge field structure function. As we can see, the right hand sides of these equations contain both 
the $\beta$ and $\gamma$ functions. 
Since we only have the asymptotic expansions of 
these two functions in the limit $g^2\ra0$, to obtain the $k^2\ra\infty$ limit of $R$ or $\ab$, we need to have  $\gb^2\ra0$ for $u\ra\infty$. 
In order to make this condition hold true, we make two basic assumptions that are  equivalent to the requirement of asymptotic freedom. First, we 
require $\beta_0 < 0$, which is equivalent to $N_F \leq 16$ in QCD. 
Second, we assume that $g^2>0$ is such that for any $0<\gb^2\leq g^2$,
$\beta(\gb^2)\neq0$. 
Given these two assumptions, one can show that $\gb^2(u,g^2) \ra +0$ for $u\ra+\infty$
with the approach
\be
\gb^2(u,g^2)\simeq (-\beta_0 \ln u)^{-1} + \cdots.
\label{gu}
\en

Since we always choose $\ka^2<0$, 
the variable $u$ is positive when $k^2<0$. Hence 
the limit $\gb^2\ra+0$,
 corresponds to the limit $k^2\ra-\infty$ along the negative real axis.

\subsection{Asymptotic limits}

In this section we give a brief overview of the possible asymptotic limits for the structure function $R(\gb^2;g^2,\al)$ for the gauge field propagator in the limit $\gb^2\ra+0$, corresponding to $k^2\ra-\infty$ along the negative real $k^2-$axis. The function $R$ satisfies the renormalization group equation (\ref{16}). It may be written as Eq.(\ref{abar}), 
which is sometimes more convenient.

Since we impose the normalization 
\be
R(g^2;g^2,\al) = 1,
\label{norm2}
\ee
we can obtain the integral equation
\be
R(\gb^2;g^2,\al) = \exp\int^{\gb^2}_{g^2}dx\frac{\ga(x;\ab(x;g^2,\al))}{\ba(x)}.
\label{Rint}
\ee
The renormalization group function $\beta(g^2)$ and the anomalous dimension $\ga(g^2,\al)$ are known as asymptotic expansions for $g^2\ra+0$. The expansion of the integrand in Eq.(\ref{Rint}) is given by
\be
\frac{\ga(x,\ab)}{\ba(x)}
\simeq \frac{\ga_{00}+\ga_{01}\ab}{\beta_0x} +
    \left(\frac{\ga_{1}(\ab)}{\beta_0}
        -\frac{\ga_{0}(\ab)\beta_1}{\beta_0^2}\right)
    +\cdots.
\label{gabeR}
\en

It is of interest to first write down several cases where the function $R$ can be obtained exactly in terms of $\ga$ and $\beta$ or truncations of $\ga/\beta$.

For $\al=0$, we have the solution (\ref{zeroa}), 
and with the expansion (\ref{gabeR}), we can write 
\be
R(\gb^2;g^2,0) = C_R(g^2,0)\left(\gb^2\right)^{\frac{\ga_{00}}{\beta_0}}
+\cdots,
\ee
where $C_R$ given in (\ref{CR}).
It is important that $C_R(g^2,0)>0$. The asymptotic limits are given by 
\be
\lim_{\gb^2\ra+0}R(\gb^2;g^2,0) = \left\{
    \begin{array}{ll}
    0 & \mbox{for $\frac{\ga_{00}}{\beta_0}>0$}\\
    +\infty & \mbox{for $\frac{\ga_{00}}{\beta_0}<0$}
    \end{array} \right.
\label{Rlimit0}
\ee
The special case $\ga_{00}=0$ will be considered later.

In the linear $\al-$approximation, where we write 
\be
\ga(g^2,\al) = \bar{\ga_0}(g^2)+\al\bar{\ga_1}(g^2) +O(\al^2),
\ee
 we can solve Eq.(\ref{16}) exactly. With the normalization (\ref{norm2}), we obtain 
\be
R(\gb^2;g^2,\al)&=&\exp\left(\int^{\gb^2}_{g^2}dx\frac{\bar{\ga}(x)}{\ba(x)}\right)\cr
    & &\times\left\{1+\al\exp\int^{\gb^2}_{g^2}dx\frac{\bar{\ga_1}(x)}{\ba(x)}
    \exp\left(\int^{x}_{g^2}dy\frac{\bar{\ga_0}(y)}{\ba(y)}\right)\right\},
\ee
and the leading asymptotic terms for $\gb^2\ra+0$ become in this approximation
\be
R(\gb^2;g^2,\al)\simeq \frac{\al}{\az} +  C_R(g^2,\al)\left(\gb^2\right)^{\frac{\ga_{00}}{\beta_0}}+\cdots,
\label{R1}
\ee
with $\az\equiv-\ga_{00}/\ga_{01}$ (in QCD, $\az=\frac{4}{9}\left(\frac{39}{4}-N_F\right)$), and 
\be
\bar{\ga_0}(g^2) = \ga_{00}g^2+O(g^4), ~~\bar{\ga_{1}}(g^2)=\ga_{01}g^2+O(g^4).
\ee
The coefficient $C_R$ is given by
\be
C_{R} (g^2,\alpha)& = & (g^2)^{-\gamma_{00}/\beta_0} \
\hbox{exp}\left( \int^0_{g^2} dx \tau_0 (x)\right)\cr
 & & \times\left\{ 1 - \frac{\alpha}{\alpha_0} + \frac{\alpha}{\alpha_0}
(g^2)^{\gamma_{00}/\beta_0} \int^0_{g^2} dx x^{-\gamma_{00}/\beta_0}
f(x,g^2)\right\}.
\label{CR1}
\en
Here
\be
\tau_0(x) &=& \frac{\gamma(x,0)}{\beta(x)}-\frac{\gamma_{00}}{\beta_0x},\cr
f (x,g^2)& =& \{ \tau_0 (x) + \alpha_0 \tau_1 (x)\}
\hbox{exp}\int^{g^2}_x dy \tau_0 (y),\cr
\tau_1(x)& =& {\bar \gamma}_1 (x)/\beta (x) - \gamma_{01}/\beta_0 x,\cr
\gamma (x,\alpha)& =& {\bar \gamma}_0 (x) + \alpha
{\bar \gamma}_1 (x) + O(\alpha^2).
\label{tau1}
\en
The expression of $C_R$ in Eq.(\ref{CR1}) reduces for $\al=0$ 
to that in Eq.(\ref{CR}).

In a one-loop approximation to $\ga/\beta$, where 
\be
\frac{\ga(g^2;\al)}{\ba(g^2)}
\approx \frac{\ga_{00}+\ga_{01}\al}{\beta_0g^2},
\en
we find the expression
\be 
R(\gb^2;g^2,\al)\approx\frac{\al}{\az} +  \left(1-\frac{\al}{\az}\right)
            \left(\frac{\gb^2}{g^2}\right)^{\frac{\ga_{00}}{\beta_0}},
\label{R1loop}
\ee
so that 
\be
C_R(g^2,\al)\approx\left(1-\frac{\al}{\az}\right)
        \left(g^2\right)^{-\frac{\ga_{00}}{\beta_0}},
\label{CR1loop}
\ee
in this approximation. Note that the coefficient $C_R$ is determined by the renormalization condition (\ref{norm2}), hence we use here the strict one-loop approximation (\ref{R1loop}) in the interval $0\leq\gb^2\leq g^2$.

In general, since $R$ is the inverse renormalization factor $Z_3=R^{-1}$, where
\be
A_\mu(x, g', \alpha', \ka') = \sqrt{Z_3}A_\mu(x, g, \alpha, \ka).
\en
and $\al'=\ab$, $g'=\gb$ as given in Eq.(\ref{4}), we want $R$ to be non-negative. This feature is also expressed by the integral equation (\ref{Rint}) as long as the integrand has no singularities in the interval $0<\gb^2\leq g^2$,
so that the exponent remains real. For $\al\neq0$, this requires that there is no zero of $R(\gb^2;g^2,\al)$ in this region $0<\gb^2\leq g^2$.

We now ask under which conditions the leading term of $R$ for $\gb^2\ra+0$ is determined by the asymptotic expressions of $\beta(\gb^2)$ and $\ga(\gb^2,\ab)$
for $\gb^2\ra0$. Since $\ab=\al/R$, this is certainly the case if $R(0)$ is finite and not zero, or if it is infinite. If the limit is finite, we see from the renormalization group equation (\ref{16}) that 
\be
R(0) = \frac{\al}{\az},
\label{Rlimit}
\ee
since we must have $\ga_{0}(\ab) = \ga_{00}+\ab\ga_{01} \ra 0$, and $\ga_0(\az)=0$. 
If the limit is infinite, it can be obtained only for $\frac{\ga_{00}}{\beta_0}<0$ with 
\be
R(\gb^2;g^2,\al)\simeq C_R(g^2,\al)\left(\gb^2\right)^{\frac{\ga_{00}}{\beta_0}}+\cdots,
\label{Rlimit-}
\ee
as seen from Eq.(\ref{16}) and (\ref{Rint}).
In situations where the one-loop term and our positivity conditions do not allow the limits (\ref{Rlimit}) and (\ref{Rlimit-}), the only other possible value for $R(0)$ is $+0$ \cite{nn}, even though here the asymptotic expansion (\ref{gabeR}) of $\ga/\beta$ is generally no more useful.

Let us now consider the cases $\ga_{00}/\beta_0>0$ and $<0$ separately:

\bigskip
\bigskip

$\ga_{00}<0$: in this case $\xi\equiv\ga_{00}/\beta_0>0$, corresponding to $N_F\leq9$ in QCD, and $\az\equiv-\ga_{00}/\ga_{01}>0$, we have always $\ga_{01}>0$ and $\beta_0<0$. Since we require $R(0)>0$, the finite limit $R(0)=\al/\az$ is possible for $\al\geq0$. With $\xi>0$, we do not have the possibility of $R(0)=+\infty$ with one-loop dominance, as can be seen from Eq.(\ref{R1}). The reason is that in this case $C_R(\gb^2)^\xi\ra0$. Summarizing the situation for $\al\geq0$, we find in the case $\ga_{00}<0$:
\be
\lim_{\gb^2\ra+0}R(\gb^2;g^2,\al) = \left\{
    \begin{array}{ll}
    \al/\az & \mbox{for $\al>0$}\\
    +0& \mbox{for $\al=0$}
    \end{array} \right.
\label{limitR}
\ee
or
\be
\lim_{\gb^2\ra+0}\ab(\gb^2;g^2,\al) = 
    \az ~~~~\mbox{for $\al>0$}.
\label{alimit}
\ee
We find $\ab\equiv0$ for $\al=0$.
If we allow values of $\al<0$, we have $\al/\az<0$, and hence the one-loop dominant limit (\ref{limitR}) is not acceptable. It also would imply a zero of $R(\gb^2;g^2,\al)$ in the interval $0<\gb^2\leq g^2$, since $R(g^2;g^2,\al)=1$. For an acceptable solution outside the realm of applicability of the asymptotic expansion (\ref{gabeR}) for the limit $g^2\ra0$, we could then have the default limit $R(0)=+\infty$. But in this paper, we consider only $\al\geq0$, where we have limits (\ref{limitR}) and (\ref{alimit}).

\bigskip
\bigskip

$\ga_{00}>0$: Here $\xi\equiv\ga_{00}/\beta_0<0$ and $\az\equiv-\ga_{00}/\ga_{01}<0$. The one-loop part of the ratio 
$\ga/\beta$ in Eq.(\ref{gabeR}) gives a leading asymptotic term $C_R(g^2,\al)(\gb^2)^{-|\xi|}$ which is acceptable for $C_R(g^2,\al)>0$.
For $\al=0$, we have the limits (\ref{Rlimit0}), again given by $C_R(g^2,0)(\gb^2)^{-|\xi|}$ with $C_R(g^2,0)>0$ as shown in Eq.(\ref{CR}). If we denote the possible zero of $C_R(g^2,\al)$ nearest to $\al=0$ by $\az(g^2)$, then, for $C_R(g^2,\az(g^2))=0$, the one-loop dominated limit is indicated. It is given by $\az(g^2)/\az$. Because of $\az<0$, it is positive for $\az(g^2)<0$. We  expect that $\az(g^2)<0$, since $C_R(g^2,0)>0$ and $\az(0)=\az<0$, as may be seen from Eq.(\ref{CR1loop}) in the strict one-loop case. For $\al<\az(g^2)$ we have $C_R(g^2,\al)<0$ and the default limit would again be $R(0)=+0$ \cite{nn}.

Summarizing, we have for 
$\ga_{00}/\beta_0<0$:
\be
\lim_{\gb^2\ra+0}R(\gb^2;g^2,\al) = \left\{
    \begin{array}{ll}
    +\infty &\mbox{for $C_R(g^2,\al)>0$; $\al>\az(g^2)<0$}\\
    +\infty &\mbox{for $\al=0$, since $C_R(g^2,0)>0$}
    \end{array} \right.
\ee
In terms of $\ab(\gb^2;g^2,\al)$ we get correspondingly
\be
\lim_{\gb^2\ra+0}\ab(\gb^2;g^2,\al) =   0 ~~~~\mbox{for $C_R(g^2,\al)>0$},
\ee
and $\ab\equiv0$ for $\al=0$.

We see that, for $\al\geq0$, we can rely upon the asymptotic expansion of $\ga$ and $\beta$ functions for $g^2\ra0$ in all cases. In the following, we consider the non-negative values of $\al$ only.

\subsection{Asymptotic expansion and sum rules}

In the case $\alpha \neq 0$, it is convenient to derive first the asymptotic 
expansion of $\ab(\gb^2;g^2,\al)$ for $\gb^2\ra0$. The function $\ab(\gb^2)$
satisfies Eq.(\ref{abar})
\be
\frac{\partial \ab(\gb^2;g^2,\al)}{\partial\gb^2} = \frac{-\ab\ga(\gb^2,\bar\al)}{\ba(\gb^2)}.
\label{abar2}
\en
Because 
the right hand side is singular at $\gb^2=0$, the equation has
a singular solution 
that is non-unique for a given initial value 
at $\gb^2=0$.  In fact, keeping 
only the one-loop term in Eq.(\ref{abar2}), we get $\ab \simeq \az + C(\gb^2)^\xi$ for $\xi>0$ and $\ab\simeq C(\gb^2)^{-\xi}$ for $\xi<0$. These leading 
term solutions suggest that $\ab$ has a branch point singularity at $\gb^2=0$. 
In order to solve Eq.(\ref{abar2}), we  introduce new
variables that uniformize the branch point at $\gb^2=0$.
Given that $N_F$ is an integer, $\xi\equiv \ro/\beta_0$ is an algebraic
number. Hence $|\xi| = n/d$ with  $n$ and $d$ being positive 
integers that are relative prime to each other. 
Because we assume $\beta_0<0$, $\xi>0$ also implies that $\xi<1$, or $n<d$.
Then, for the case $0<\xi<1$, we define
new variables:
\be
    x&\equiv& (\gb)^{\frac{1}{d}}, \nonumber\\
    y&\equiv& \f{\ab-\az}{x^n},
\label{ydef}
\ee
and for the case $\xi<0$
\be
    x&\equiv& (\gb)^{\frac{1}{d}}, \cr
    v&\equiv& \f{\ab}{x^n}.
\label{zdef}
\ee 
 
Applying the chain rule, we have:
\begin{eqnarray*}
    \f{dy}{dx} &=& \f{1}{x^n}\f{d\gb}{dx}\f{d\ab}{d\gb} 
            - \f{n(\ab-\az)}{x^{n+1}}  \\
        &=& -d\,x^{d-n-1}\left( \f{\ab\gamma(\ab,\gb)}{\beta(\gb)}
         - \f{\az\gamma_0(\ab)}{\beta_0 x^d} \right)  \\
        &=& -d\,x^{d-n-1}\left( \f{\ab\gamma(\ab,\gb^2)}{\beta(\gb^2)} 
             - \f{\ab\gamma_0(\ab)}{\beta_0x^d} +
            \f{\ab\gamma_0(\ab)}{\beta_0x^d} 
            -\f{\az\gamma_0(\ab)}{\beta_0 x^d}\right) \\
        &=& -d\,x^{d-n-1}\ab\left( \f{\gamma(\ab,\gb^2)}{\beta(\gb^2)} 
          - \f{\gamma_0(\ab)}{\beta_0x^d}\right) 
          -  d\,x^{d-n-1}(\ab-\az)\f{\gamma_0(\ab)}{\beta_0 x^d} 
\end{eqnarray*}
Substituting $\ab = \az+x^ny$, we get
\be
\frac{dy}{dx} = \frac{n}{\alpha_0} x^{n-1} y^2 - d~ x^{d-n-1} (\alpha_0
 + x^n y) \phi (x^d,\alpha_0  + x^n y),
\label{yx}
 \en
 where
 \begin{eqnarray}
 \phi (g^2,\alpha)&=&\frac{\gamma (g^2,\alpha)}{\beta (g^2)} -
 \frac{\gamma_0 (\alpha)}{\beta_0 g^2} \simeq\phi_0 (\alpha) + g^2\phi_1
 (\alpha) + \cdots\cr
 \phi_0 (\alpha)& =& \frac{\gamma_1 (\alpha)}{\beta_0} -
 \frac{\beta_1}{\beta_0^2} \gamma_0 (\alpha),~~~\hbox{etc.}.
\label{phi}
\end{eqnarray}
In the first term on the right hand side of (\ref{yx}),
$n-1\geq0$ because $n$ is a positive integer. In the second
 term, $d-n-1\geq0$ because $n<d$. Furthermore,
in an appropriate finite domain including $g^2=0$, and excluding
 possible, nontrivial fixed points corresponding to zeroes of $\beta
 (g^2)$, it is reasonable to assume that $\phi (g^2,\alpha)\simeq\phi_0 (\alpha) + g^2\phi_1
 (\alpha) + \cdots$ is
 continuously differentiable.  As far as $\beta (g^2)$ and $\gamma
 (g^2,\alpha)$ are represented by power series expansions for $g^2\to +
 0, \phi (g^2,\alpha)$ is also a power series in $g^2$ and $\alpha$.
 Under these circumstances, the right hand side of
  (\ref{yx}) satisfies the Lipschitz
 condition for $x=+0$, 
i.e. 
there exists a constant $L$, such that for any $x$ in the interval
of $[0,\epsilon)$, and any $y'$ and $y''$,
\be
\left|\frac{dy}{dx}(x,y') - \frac{dy}{dx}(x,y'')\right|\leq
L\left|y'-y''\right|.
\en
We then have exactly one solution through every
 point $x=0, y=C$.  In as far as the right hand side
 of (\ref{yx}) is also a power
 series, we obtain the solution in the form of a series.
Thus, for any finite constant $C$, there is a unique solution $y(x)$ such that $y(0)=C$. Substituting $y = C + \sum_{j=1}^\infty a_jx^j$, we obtain for the solution of Eq.(\ref{yx})
\be
 y(x) &=& C + \frac{C^2}{\alpha_0} x^n
  +C \left( \frac{\xi + 1}{\xi - 1} \phi_0 (\alpha_0) - \alpha_0 \phi'_0
 (\alpha_0)\right) x^d + \cdots \nonumber\\
    & & + \frac{\alpha_0}{\xi - 1} \phi_0 (\alpha_0) x^{d-n} + \cdots.
\label{ysoln}
\ee

The power series in Eq.(\ref{ysoln}) gives the asymptotic expansion of $y$ for $x\ra+0$.
By definition (\ref{ydef}), we have,
for $0<\xi<1$, and $\gb^2\ra+0$:
\be
 \ab &\simeq& \az + C(\gb^2)^\xi + \frac{C^2}{\alpha_0} (\gb^2)^{2\xi}
  +C \left( \frac{\xi + 1}{\xi - 1} \phi_0 (\alpha_0) - \alpha_0 \phi'_0
 (\alpha_0)\right) \gb^2 + \cdots \nonumber\\
    & & + \frac{\alpha_0}{\xi - 1} \phi_0 (\alpha_0) (\gb^2)^{1-\xi} + \cdots,
\label{absoln}
\ee
and $C = C(g^2,\al) = \lim_{\gb^2\ra0}\left(
\ab(\gb^2;g^2,\al)-\az\right)(\gb^2)^{-\xi}$. The coefficient $C(g^2,\al)$ contains all the gauge dependence of the above expansion.

The expansion in Eq.(\ref{absoln}) is valid if and only if the constant $C(g^2,\al)$ is finite. In order for this to be true, it is necessary, with the form of $C(g^2,\al)$ given above, that $\ab\ra\az$ for $\gb^2\ra0$. It can be shown
 that this is also sufficient.
Then $C(g^2,\al)$ is finite if and only if $\ab\ra\az$. From the discussions given in the previous section, we can then conclude that the expansion in Eq.(\ref{absoln}) is valid for $\al>0$.

For $\al=0$, we have the trivial solution $\ab\equiv0$.

\bigskip

In the case $\xi<0$, we obtain from the definition (\ref{zdef}) that
\be \f{dv}{dx} =  \f{n}{\az}x^{n-1}v^2 -d\,x^{d-1}v\phi(x^nv,x^d).  \label{zx}
\ee
By an analogous analysis as for Eq.(\ref{yx}), Eq.(\ref{zx}) is also Lipschitz in  an interval $[+0,\epsilon)$. For any finite constant $C$, it has the unique solution  
\be
v = Cx + \f{C^2 \gamma_{01}}{\ro}x^n + \f{C}{\beta_0}(\gamma_{10} - \beta_1\xi)x^{d+1} + \cdots. 
\ee
By definition (\ref{zdef}), we have,
for $\xi<0$, and $\gb^2\ra0$:
\be
\ab \simeq C(\gb^2)^\xi + \f{C^2 \gamma_{01}}{\ro}(\gb^2)^{2\xi}
        + \f{C}{\beta_0}(\gamma_{10} - \beta_1\xi)(\gb^2)^{1+\xi} + \cdots.
\label{absoln2}
\ee
In this case, $C(g^2,\al) = \lim_{\gb^2\ra0}\ab(\gb^2;g^2,\al)(\gb^2)^{-\xi}$, 
and in general it is a different function from $C(g^2,\al)$ in Eq.(\ref{absoln}).  Again the constant $C(g^2,\al)$ contains all the gauge dependence of the expansion (\ref{absoln2}).

\bigskip
\bigskip

Since we assumed $k^2<0$, we can relate the $k^2\ra-\infty$ limits to those for $\gb^2\ra+0$ via Eq.(\ref{gu}). Then from 
$-k^2D(k^2,g^2,\al,\ka^2)=R(\frac{k^2}{\ka^2},g^2,\al)=\al\ab^{-1}(\gb^2;g^2,\al)$, and the expression given in (\ref{zeroa}) for the case $\al=0$,
we obtain the following expansions for $\al\geq0$
in the limit $k^2\ra -\infty$:

If $0<\frac{\ga_{00}}{\beta_0}<1$ ($N_F\leq9$ in QCD with $\beta_0<0$):
\be
-k^2D(k^2, g^2,\alpha,\ka^2) &\simeq& \frac{\alpha}{\alpha_0} + C_{R} \left(-\beta_0\ln \frac{k^2}{\ka^2}\right)^{-\frac{\gamma_{00}}{\beta_0}} \nonumber\\
& & +C_{R} \beta^{-1}_0 \left(\gamma_{10} - \gamma_{12}\alpha_0^2 -
 \frac{\beta_1}{\beta_0}\gamma_{00}\right) \left(-\beta_0\ln \frac{k^2}{\ka^2}\right)^{-\frac{\gamma_{00}}{\beta_0} - 1} +\cdots\nonumber\\
    & &+ \frac{\alpha}{\alpha_0} \frac{\gamma_1 (\alpha_0)}{\beta_0-\gamma_{00}}
 \left(-\beta_0\ln \frac{k^2}{\ka^2}\right)^{-1}  + \cdots,
\label{d+}
\ee

If $\frac{\ga_{00}}{\beta_0}<0$ ($10\leq N_F\leq16$ in QCD with $\beta_0<0$):
\begin{eqnarray}
  -k^2D(k^2, g^2,\alpha,\ka^2)&\simeq&C_{R}\left(-\beta_0\ln \frac{k^2}{\ka^2}\right)^{-\frac{\gamma_{00}}{\beta_0}}\nonumber\\
    & &  + C_{R}\frac{1}{\beta_0} (\gamma_{10} - \frac{\beta_1}{\beta_0} 
    \gamma_{00})
    \left(-\beta_0\ln \frac{k^2}{\ka^2}\right)^{-\frac{\gamma_{00}}{\beta_0}- 1}+\cdots\cr
    & &+ \frac{\alpha}{\alpha_0} +
 \frac{\alpha}{\alpha_0} \frac{(\gamma_{10} +
 \alpha_0 \gamma_{11})}{\beta_0-\gamma_{00}} \left(-\beta_0\ln \frac{k^2}{\ka^2}\right)^{-1} + \cdots.\nonumber\\
\label{d-}
 \end{eqnarray}
In the above, $C_{R} \equiv - C\alpha/\alpha^2_0$, and is not necessarily equal in the two cases.

An important aspect of the asymptotic expansions in Eq.(\ref{d+}) and Eq.(\ref{d-}) is that all the asymptotic terms are gauge independent, except for the coefficient $C_R(g^2,\al)$. Furthermore, the orders of all the asymptotic terms are exactly determined by the one-loop coefficients $\gamma_{00}$ and $\beta_0$ of the anomalous dimension $\gamma$ and the renormalization group function $\beta$.

A priori, the coefficients $C_{R}$
 appearing in the solutions of
 the nonlinear, ordinary differential equations are undetermined
 constants. However, because of the normalization condition $R(g^2;
 g^2,\alpha) =1$ or $\bar\alpha (g^2; g^2,\alpha) = \alpha$, the
 coefficients become functions of $g^2$ and $\alpha$, satisfying
 partial differential equations in these variables.  For $C_{R}
 (g^2,\alpha)$, we find the equation
 \begin{eqnarray}
 C_{R} (g^2,\alpha) = R(g'^{2}; g^2,\alpha) C_{R} (g'^{2},\alpha')
  \end{eqnarray}
with $g'$ and $\al'$ given in Eq.(\ref{4}).
The corresponding differential equation for $C_R$ is:
\be
 \beta (g^2) \frac{\partial C_{R}}{\partial g^2} = \alpha \gamma
 (g^2,\alpha) \frac{\partial C_{R}}{\partial\alpha} - \gamma
 (g^2,\alpha) C_{R}.
\label{cgg}
\en

 For $\alpha = 0$, and in the $\alpha$--linear approximation, we have
 given $C_{R}$ in (\ref{CR}) and (\ref{CR1}), which satisfy
  (\ref{cgg}) with $\alpha = 0$ and $\ga (g^2, \alpha) =
{\bar \ga}_0 (g^2) + \alpha {\bar \ga}_1 (g^2)$ respectively.

\bigskip
\bigskip

So far, we have obtained the asymptotic expansion of the gluon propagator 
for the limit $k^2\ra-\infty$ along the negative real axis.
To extend these results to 
the limits of $k^2\ra\infty$ along all directions in the complex $k^2$-plane, we return to the renormalization group transformation of the dimensionless structure function, Eq.(\ref{7}). Setting $\ka^{\prime 2} = - \vert
 k^2\vert$, we find, with $\ka^2 < 0$ and $k^2 = - \vert k^2\vert
 e^{i\varphi}$ for all $ \vert\varphi\vert \leq \pi$:
 \be
R \left( \frac{k^2}{\ka^2},g,\alpha\right) =
 R \left(\left\vert \frac{k^2}{\ka^2}
 \right\vert, g,\alpha\right) R(e^{i\varphi},\bar g, \bar\alpha).
 \label{rep}
 \en
 Here $\bar g = \bar g (\vert\frac{k^2}{\ka^2}\vert,g)$ and $\bar\alpha =
 \alpha R^{-1} (\vert\frac{k^2}{\ka^2}\vert,g,\alpha)$.
  For $\beta_0 < 0$, the effective coupling $\bar g^2$
 vanishes for $\vert\frac{k^2}{\ka^2}\vert\to\infty$, and  $\bar\alpha$
 approaches to finite limit for $\al\geq0$.  Because 
the function $R$ is analytic in the cut
 complex $k^2$--plane, we can then use the perturbation expansion for the
 structure function,
 \be
R \left( \frac{k^2}{\ka^2},g,\alpha \right) \simeq 1 + g^2
 \gamma_0 (\alpha ) \ln \left ( \frac{k^2}{\ka^2} \right ) + O(g^4),
 \en
 and write for $\bar g^2\to + 0$:
 \be
R (e^{i\varphi},\bar g,\bar\alpha)\simeq 1+\bar g^2
 \gamma_0 (\bar \alpha) i\varphi + O({\bar g}^4) .
\label{rph}
 \en
 Eq.(\ref{rep}) expresses
the asymptotic limit for $k^2\to\infty$ in
 all directions in terms of the limit along the negative real
 $k^2$--axis.  With Eqs.(\ref{rep}), (\ref{rph}), (\ref{d+}) and 
(\ref{d-}), we find that the 
asymptotic expansions in  
 Eq.(\ref{d+}) and (\ref{d-}) are also valid in all directions in the 
$k^2-$plane.

>From Eq.(\ref{d+}) and Eq.(\ref{d-}), we can derive
 the discontinuity of the structure function $-k^2D$ across the positive real 
$k^2-$axis. Let 
\be
2i\pi\rho(k^2)\equiv \lim{\epsilon\ra0}\left(D(k^2+i\epsilon)-D(k^2-i\epsilon)\right).
\ee 
It is straightforward to derive that for $\ka^2<0$ the discontinuity of the function $(\ln\frac{k^2}{\ka^2})^A$ at real positive $k^2$ is given by $-A(\ln\frac{k^2}{|\ka|^2})^{A-1}$. We then have:

\bigskip

\noindent for $0<\xi<1$,
\be
-k^2\rho(k^2, g^2,\alpha,\ka^2) &\simeq& \frac{\gamma_{00}}{\beta_0}
    C_{R} \left(-\beta_0\ln \frac{k^2}{|\ka|^2}\right)
    ^{-\frac{\gamma_{00}}{\beta_0}-1} \nonumber\\
    & & +C_{R} \frac{\gamma_{00}+\beta_0}{\beta_0^2}
     \left(\gamma_{10} - \gamma_{12}\alpha_0^2 -
    \frac{\beta_1}{\beta_0}\gamma_{00}\right) 
    \left(-\beta_0\ln \frac{k^2}{|\ka|^2}\right)
        ^{-\frac{\gamma_{00}}{\beta_0} - 2}\cr
    & &+\cdots\nonumber\\
    & &+ \frac{\alpha}{\alpha_0} \frac{\gamma_1 (\alpha_0)}{\beta_0-\gamma_{00}}
    \left(-\beta_0\ln \frac{k^2}{|\ka|^2}\right)^{-2}  + \cdots
\label{rho+}
\ee

\noindent for $\xi<0$,
\begin{eqnarray}
  -k^2\rho(k^2, g^2,\alpha,\ka^2)&\simeq&
    \frac{\gamma_{00}}{\beta_0}
    C_{R}\left(-\beta_0\ln \frac{k^2}{|\ka|^2}\right)
        ^{-\frac{\gamma_{00}}{\beta_0}-1}\nonumber\\
    & &  + C_{R}\frac{\gamma_{00}+\beta_0}{\beta_0^2} 
    (\gamma_{10} - \frac{\beta_1}{\beta_0} 
    \gamma_{00})
    \left(-\beta_0\ln \frac{k^2}{|\ka|^2}\right)
        ^{-\frac{\gamma_{00}}{\beta_0} - 1}+\cdots\cr
    & &+ \frac{\alpha}{\alpha_0} 
    \frac{(\gamma_{10} +\alpha_0 \gamma_{11})}{\beta_0-\gamma_{00}} 
    \left(-\beta_0\ln \frac{k^2}{|\ka|^2}\right)^{-2} + \cdots.\nonumber\\
\label{rho-}
 \end{eqnarray}

The solutions in Eq.(\ref{d+}) and (\ref{d-}) also imply certain exact relations along the positive real axis of the $k^2-$plane. Because $D(k^2)$ is analytic in the cut $k^2-$plane, the Cauchy's theorem implies that
\be
D (k^2,\ka^2,g,\alpha) &=& \oint\f{D(k'^{2})}{k'^{2}-k^{2}}dk'^{2}+
    \int^{\Delta}_{-0} dk'^{2}  \frac{\rho (k'^{2},
 \ka^2, g, \alpha)}{k'^{2} - k^2}
\ee
where the first integral is made along a circle in the complex $k^2-$plane
with radius $\Delta$.

In both Eq.(\ref{d+}) and (\ref{d-}), 
$D(k^2,\ka^2,g^2,\al)$ vanishes as $k^2\ra\infty$ along all directions in the complex $k^2-$plane. The Cauchy's theorem then implies the unsubtracted dispersion relation
\be
D(k^2,\ka^2,g^2,\al) = \int_{-0}^\infty dk'^2\frac{\rho(k'^2,\ka^2,g^2,\al)}{k'^2-k^2}.
\label{dispersion}\ee

We also have sufficient boundedness for the discontinuity $\rho$ in Eq.(\ref{rho+}) and (\ref{rho-}) to write a dipole representation
\be
D(k^2,\ka^2,g^2,\al) &=& \int_{-0}^\infty dk'^2\frac{\sigma
 (k'^{2},\ka^2,g,\alpha)}{(k'^{2} - k^2)^2},
\ee
where
\be
 \sigma (k^2,\ka^2,g,\alpha) &\equiv& \int^{k^2}_{-0} dk'^{2} \rho
 (k'^{2},\ka^2,g,\alpha).
 \end{eqnarray}
For $\alpha = 0$, this dipole representation has been discussed
 in  \cite{126, Np} in connection with an approximately linear
 quark--antiquark potential.

 Of particular interest is the situation for $0<\gamma_{00}/\beta_0 < 1$,
 corresponding to $N_F\le 9$ in QCD.  There, the function $-k^2D \ra
 \frac{\alpha}{\alpha_0}$ for
 $k^2\to\infty$ in all directions, and hence we obtain from 
the unsubtracted dispersion relation Eq.(\ref{dispersion})
the sum rule
 \be
\int^\infty_{-0} dk^2\rho (k^2,\ka^2,g,\alpha) &=&
 \frac{\alpha}{\alpha_0}.
\label{sup}
 \en
 This is the generalization of the superconvergence relation
\cite{oz,128}
 $$\int^\infty_{-0} d k^2 \rho (k^2,\ka^2,g,0) ~~=~~ 0,
 $$
 which was obtained previously in the Landau gauge.  The relation
(\ref{sup}) expresses the fact that the coefficient of the
asymptotic term proportional to $k^{-2}$ in the representation
(\ref{dispersion}) is given by $-\alpha / \alpha_0 $. It
is {\it not} valid for $\gamma_{00}/\beta_0 < 0$.
The distribution aspects of sum rules like (\ref{sup}),
and of the related dispersion representations, have been discussed
in \cite{oz, 128}.

\bigskip

So far, we have not considered the case $\gamma_{00}=0$, because in 
ordinary gauge theory we have $\ga_{00}=0$ for 
$N_F=\frac{13}{4}N_C$, so in QCD with $N_C=3$, it is not of direct interest.
 But 
it could be realized for $N_C=4$, etc. In $N=1$ SUSY SU($N_C$) 
theories with matter 
fields in the regular representation, we have $\ga_{00}=0$ 
(Wess-Zumino representation) 
for $N_F=\frac{3}{2}N_C$, and it is of interest for $N_C=$ even. 
For completeness, we give in the following  the 
large $k^2$ expansion of the gauge structure function for 
$\ga_{00}=0$.

For $\al=0$, the gauge structure function has been given in Eq.(\ref{zeroa}).
With $\xi=0$, we then have for $k^2\ra\infty$,
\be
-k^2D(k^2) &\simeq& R(0;g^2,0)\left\{1+\frac{\ga_{10}}{\ba_0}\left(-\ba_0
\ln \frac{k^2}{\kappa^2} \right)^{-1} + \cdots \right\} ,
\label{35}
\en
with $R(0;g^2,0) = \exp\int^{0}_{g^2} dx \frac{\ga(x,0)}{\ba(x)}$ which is finite. There is 
the sum rule
\be
\int^{\infty}_{-0} dk^2 \rho(k^2,\ka^2,g,0) = R(0;g^2,0) .
\label{36}
\en
For $\al>0$, the leading asymptotic term is  
\be
-k^2D(k^2) &\simeq& -\frac{\ga_{01}\al}{\beta_0}\ln\ln\frac{k^2}{\ka^2}
    +C_{R0}(g^2,\al)+\cdots ,
\label{D0}
\en
where $C_{R0}$ is a finite constant such that $C_{R0}(g^2,0)=R(0;g^2,\al)$. 
There is no sum rule in this case.
The leading term of the discontinuity is given by
\be
-k^2\rho(k^2) &\simeq& -\frac{\ga_{01}\al}{\beta_0}\left(\ln\frac{k^2}
{|\ka^2|}\right)^{-1}+\cdots .
\label{39}
\en

\section{Quark Propagator}

In this section, we present general results for the quark 
propagator \cite{oxq}. We obtain 
the asymptotic terms of the structure functions in general, linear, covariant 
gauges. Except for the case of the Landau gauge, 
which has been considered before \cite{gw,oz}, the previous results for the 
gauge field propagator play an important r\^{o}le in the derivation. 
In fact, the quark structure function can be expressed 
in a closed form in terms of the gauge field structure function. Again,
in our results, only 
anomalous dimension and renormalization group function 
coefficients in lowest non-vanishing 
order of perturbation theory are relevant, and gauge dependence is limited.

Recall that the ``quark'' propagator 
\be
S_{F}(k^2+i0) = \int d^4x \exp^{ik\cdot x}
\langle 0|T\psi(x)\bar{\psi}(0)|0\rangle 
\label{1}
\en
can be written in the form
\be
S_F(k^2+i0) = A(k^2+i0)\sqrt{-k^2} + B(k^2+i0)\ga\cdot k ~,
\label{2}
\en
where gauge and spinor indices have been suppressed.
The two components of the quark propagator, $A$ and $B$,  will be shown to have different
asymptotic behavior at infinite momentum and obey different sum rules.

It is convenient to introduce dimensionless structure functions 
\be
S\left(\frac{k^2}{\ka^2},g,\al\right)=-k^2A(k^2,\ka^2,g,\al),\cr
T\left(\frac{k^2}{\ka^2},g,\al\right)=-k^2B(k^2,\ka^2,g,\al).
\label{5}
\en
with the normalization
\be
T(1,g,\al) = 1 .
\label{6}
\en
With $\ka^2 < 0$ as the normalization point, we have the renormalization group equation
\be
\psi(x, g',\al',\ka'^2) = \sqrt{Z_2}\psi(x,g,\al,\ka^2)~,
\label{3}
\en
and corresponding relations for other fields.
Here 
\be
Z_2 &=& Z_2\left(\frac{\ka'^2}{\ka^2},g,\al\right),
\en
and $g'$ and $\al'$ are given in Eq.(\ref{4}).
Eq.(\ref{1}) and (\ref{3}) imply that
\be
T\left(\frac{k^2}{\ka^2},g,\al\right) = Z_2^{-1}\left(\frac{\ka'^2}{\ka^2}, g, \al\right)T\left(\frac{k^2}{\ka'^2},g',\al'\right) .
\label{Z2}
\en
Setting $k^2=\ka'^2$, the normalization condition Eq.(\ref{9}) implies 
that $Z_2^{-1} = T$.
Substituting this result back to Eq.(\ref{Z2}), we get 
the renormalization group equation in terms of the dimensionless functions:
\be
T\left(\frac{k^2}{\ka^2},g,\al\right) &=& T\left(\frac{\ka'^2}
{\ka^2},g,\al\right)T\left(\frac{k^2}{\ka'^2},g',\al'\right) ,
\label{7q}
\en
\be
S\left(\frac{k^2}{\ka^2},g,\al\right) &=& T\left(\frac{\ka'^2}
{\ka^2},g,\al\right)S\left(\frac{k^2}{\ka'^2},g',\al'\right) ,
\label{8}  
\en

In a situation, where the theory has unbroken chiral symmetry, the structure 
function $A$ and $S$ vanish identically. But if we allow for possible non-perturbative 
mass generation, we expect a nonzero function, which vanishes in the perturbative limit.

As a consequence of Lorentz covariance and simple spectral conditions 
formulated in the state space of indefinite metric, it follows that the 
structure functions (distributions) $B(k^2+i0)$ and $A(k^2+i0)$ are boundary values of 
corresponding analytic functions, which are regular in the cut $k^2$-plane 
with branch lines along the positive real $k^2$-axis. In the following, 
we first use the renormalization group in order to obtain the asymptotic 
behavior of these functions for $k^2\rightarrow -\infty$ along the negative 
real $k^2-$axis. We then generalize the results to all directions in the 
complex $k^2$-plane \cite{ox,oz}. For this purpose, we consider Eq.(\ref{7q})
and (\ref{8}) with 
$\ka'^2=-|k^2|$, $k^2=-|k^2|e^{i\phi}, |\phi| \leq \pi$, and find 
\be
T\left(\frac{k^2}{\ka^2},g,\al\right) = T\left(\left|\frac{k^2}{\ka^2}\right|,g,\al\right)
T(e^{i\phi},\bar g,\bar{\al}), \\
\label{9}
S\left(\frac{k^2}{\ka^2},g,\al\right) = T\left(\left|\frac{k^2}{\ka^2}\right|,g,\al\right)
S(e^{i\phi},\bar g,\bar{\al}),
\en
where $\gb = \gb(|\frac{k^2}{\ka^2}|,g)$, and $\bar\al = \bar\al(|\frac{k^2}{\ka^2}|,g,\al)$.
In the following, we
first discuss the function $T$ only.

Assuming that the exact structure functions approach their perturbative 
limits for $g^2\rightarrow0$, at least as far as the leading terms are concerned,
we have for vanishing $g^2$:
\begin{eqnarray}
T\left(\frac{k^2}{\ka^2},g,\al\right) &\simeq& 1+g^2\ga_{F01}
\al\ln\frac{k^2}{\ka^2}+\cdots\nonumber\\
& &\hspace{2in}\mbox{for}~~ \al \neq 0\nonumber\\
T\left(\frac{k^2}{\ka^2},g,\al\right) &\simeq& 1+g^4\ga_{F01}
\ln\frac{k^2}{\ka^2}+\cdots\nonumber\\
& &\hspace{2in}\mbox{for}~~ \al = 0
\label{10}
\end{eqnarray}
In these equations, 
$\ga_{F01}$ is the coefficient from the quark
anomalous dimension defined by $\ga_F(g^2,\al)\equiv u\frac{\partial T(u,g,\al)}{\partial u}|_{u=1}$. The quark anomalous dimension can be expanded 
in perturbation theory for vanishing $g^2$ as
\be
\ga_F(g^2,\al)&\simeq&\ga_{F0}(\al)g^2+\ga_{F1}(\al)g^4 + \cdots ,
\label{gammaF}
\en
 with 
\be
\ga_{F0}(\al)&=&\ga_{F00}+\al\ga_{F01},\\
 \ga_{F1}(\al)&=&\ga_{F10}+\al\ga_{F11}+\al^2\ga_{F12}. 
\en
We consider here QCD, or similar theories, so that we always
have $\ga_{F00}\equiv 0$ and $(16\pi^2)\ga_{F01}=C_2(R)>0$, with $C_2(R)=4/3$
in QCD. For SUSY theories, on the other hand, we generally
find $\ga_{F00}\neq 0$ for the Fermi field in the Wess-Zumino representation
\cite{of}.
These theories will be discussed elsewhere.

Given asymptotic freedom, we know that
 $\gb^2(u,g)\simeq(-\ba_0\ln u)^{-1}$ and hence vanishes for $u\ra\infty$.
Furthermore, from the results we derived in the previous section,
$\bar{\al}(u,g,\al)=\al R^{-1}(u,g,\al)$
converges to finite limits $0$ or $\al_0 = -\ga_{00}/\ga_{01}$ for $u\rightarrow\infty$, at least in the case $\al \geq 0 $.
Then in view of the analytic properties of the structure functions, 
we may use the perturbation series in Eq.(\ref{10}) to evaluate the
the function $T(e^{i\phi},\bar g,\bar{\al})$ in the limit $u\ra\infty$,
corresponding to $\gb^2\ra+0$.

In the case $\al>0$, $\xi>0$,
\be
T(e^{i\phi},\bar g,\ab)\simeq 1+\gb^2\ga_{F01}\az i\phi 
            +\cdots,
\label{Tphi+}
\en
and the case $\al\geq0$, $\xi<0$, 
\be
T(e^{i\phi},\bar g,\ab)\simeq 1+\gb^4\ga_{F01}i\phi 
            +\cdots.
\label{Tphi-}
\en

In Eq.(\ref{9}), the function 
$T\left(\left|\frac{k^2}{\ka^2}\right|,g,\al\right)$ can be calculated by
 $k^2<0$. Since $\ka^2<0$, we have 
$T\left(\left|\frac{k^2}{\ka^2}\right|,g,\al\right) = 
T\left(\frac{k^2}{\ka^2},g,\al\right)$ for $k^2<0$. This function 
satisfies the renormalization
group equation Eq.(\ref{7q}).
Eq.(\ref{9}) then gives the asymptotic behavior in all directions 
$\phi$ in terms 
of that for real $k^2\rightarrow-\infty$.

To obtain the $k^2\ra-\infty$ asymptotic expansion of the
function $T$, we first transform Eq.(\ref{7q}) into a differential
equation. 
By differentiating Eq.(\ref{7q})
with respect to $k^2$, setting $\ka'^2 = k^2$, we obtain the equation
\be
u\frac{\partial T(u,g,\al)}{\partial u} = \ga_F(\gb^2,\bar\al)T(u,g,\al) ,
\label{11}
\en
where $u = k^2/\ka^2,~ k^2\leq0$ and $\gb^2 = \gb^2(u,g)$, $\bar\al = \bar\al(u,g,\al) = 
\al R^{-1}(u,g,\al)$ and $\ga_F(g^2,\al)\equiv u\frac{\partial T(u,g,\al)}{\partial u}|_{u=1}$,
$\ba(g^2)\equiv u\frac{\partial \gb^2(u,g)}{\partial u}|_{u=1}$. It is 
again convenient 
to introduce the new variable $\gb^2(u,g)$ defined in Eq.(\ref{dgdu}).
 In terms of $\gb^2$, we write
\be
T(\gb^2;g^2,\al) = T\left(\frac{k^2}{\ka^2},g,\al\right) ,
\label{12}
\en
and obtain from Eq.(\ref{11}) the equation
\be
\frac{\partial \ln T(\gb^2;g^2,\al)}{\partial\gb^2} = \frac{\ga_F(\gb^2,\bar\al)}{\ba(\gb^2)} .
\label{13}
\en
Here we have used
$\ab(u,g,\al) = \ab(\gb^2;g^2,\al) = \al R^{-1}(\gb^2;g^2,\al)$.
If we consider $R(\gb^2;g^2,\al)$ or $\ab(\gb^2;g^2,\al)$ as given, we can write the solution in the form
\be 
T(\gb^2;g^2,\al) = \exp\int^{\gb^2}_{g^2}dx\frac{\ga_{F}(x;\ab(x;g^2,\al))}{\ba(x)},
\label{14}
\en
where use has been made of the normalization condition (\ref{6}): $T(g^2;g^2,\al) = 1$.

We are interested in the asymptotic expansion of $T(\gb^2;g^2,\al)$ 
for $\gb^2\rightarrow 0$. Because of the appearance of the gauge field structure 
function $R$ for $\al\neq0$, we have to consider separately the two cases for different 
signs of the one loop gauge-field anomalous dimension coefficient $\ga_{00}=\ga_0(\al=0)$,
and the possibility that $\gamma_{00} = 0$. For the expansion of $R(\gb^2;g^2,\al)
=R(\frac{k^2}{\ka^2},g,\al)=-k^2D(k^2,\ka^2,g,\al) $ in the limit 
$\gb^2\rightarrow0$, we have obtained in the previous section the leading terms
\be
R(\bar g^2; g^2,\alpha) &\simeq&  C_{R} (\bar
 g^2)^\xi +
 C_{R} \beta^{-1}_0 \left(\gamma_{10} - \gamma_{12}\alpha_0^2 -
 \frac{\beta_1}{\beta_0}\gamma_{00}\right) (\bar g^2)^{\xi + 1} +\cdots\cr
 &~&~~~~+ \frac{\alpha}{\alpha_0} + \frac{\alpha}{\alpha_0} \frac{\gamma_1 (\alpha_0)}{\beta_0}
 \frac{1}{1-\xi} \bar g^2 + \cdots,
\label{15}
\en
where $\alpha \geq 0$ and $\xi\equiv\ga_{00}/\ba_0 \neq 0 $ has been assumed.
The special case $\gamma_{00}=0$ will be discussed later. 
For $\al=0$, the coefficient $C_R(g^2,\al)$ is given in Eq.(\ref{CR})
and in this case we have $C_R(g^2,0) > 0$. For $\al>0$, we consider only 
parameters
$(g^2,\al)$ such that $C_R(g^2,\al)>0$, as discussed in section 2.2.

\bigskip
\bigskip

${\bf \gamma_{00}/\beta_0 < 0} $ :  Let us first consider the case 
$\xi\equiv\frac{\ga_{00}}{\ba_0} < 0$, corresponding to  $\ga_{00}>0$, $\ba_0<0$.
For QCD, 
we have $\xi = \frac{\frac{13}{2}-\frac{2}{3}
N_F}{11-\frac{2}{3}N_F}$, so that $\xi<0$ corresponds to $10\leq N_F\leq16$, 
where $|\xi|<1$ for $N_F\leq13$ and $|\xi|>1 $ for $14\leq N_F\leq16$. 
\footnote{For $N=1$ 
$SU(N_C)$ SUSY gauge theory in the Wess-Zumino representation, we have $\xi<0$ for 
$\frac{3}{2}N_C<N_F<3N_C, |\xi|<1$ for $\frac{3}{2}N_C<N_F<\frac{9}{4}N_C$ and 
$|\xi|>1$ if $\frac{9}{4}N_C<N_F<3N_C.$ Here $N_F$ is the number of flavors in 
the regular representation.}

Using Eq. (\ref{15}), we expand the integrand in Eq. (\ref{14}) as an asymptotic 
series for $\gb^2\rightarrow 0$. Since
$\ga_{F00}=0$ for QCD,  we find
\be
\frac{\ga_F(\gb^2,\ab)}{\ba(\gb^2)} &\simeq& \al\frac{\ga_{F01}}{\ba_0}
C_R^{-1}(\gb^2)^{-1-\xi}
- \al\frac{\ga_{F01}}{\ba_0}\frac{\al}{\al_0}C_R^{-2}(\gb^2)^{-1-2\xi}+\cdots
\nonumber \\ & & + \frac{\ga_{F10}}{\ba_0} + \cdots .
\label{17}
\en
With $\xi<0$, the number of singular terms depends upon the magnitude $|\xi|$. 
But the leading  term with the order $(\gb^2)^{-1-\xi}$ indicates that
the function $\frac{\ga_{00}(x,\ab(x))}{\beta(x)}$ is integrable at the origin, and hence we can write Eq.(\ref{14}) as:
\be
T(\gb^2;g^2,\al) = T(0;g^2,\al)\exp\int_{0}^{\gb^2}dx\frac{\ga_{F}(x;\ab(x;g^2,\al))}{\ba(x)}.
\label{Tint-}
\en
We can expand the integral here for $\gb^2\ra+0$ by integrating term by term the expansion given in Eq.(\ref{17}). The result gives the following expansion of the function $T(\gb^2;g^2,\al)$ for $\gb^2\ra0$:
\be
T(\gb^2;g^2,\al) &\simeq& T(0;g^2,\al)\left\{ 1 -\al\frac{\ga_{F01}}
{\ga_{00}}C_R^{-1}({\gb}^2)^{-\xi} \right. \nonumber \\
& & - \al^2 \frac{\ga_{F01}\ga_{01}}{2\ga_{00}^2}\left(
1-\frac{\ga_{F01}}{\ga_{01}}\right)
C_R^{-2}({\gb}^2)^{-2\xi}  + \cdots \nonumber \\
& & \left. + \frac{\ga_{F10}}{\ba_0}{\gb}^2 +\cdots \right\},
\label{18}
\en
where terms of order $(\gb^2)^{1-\xi}, (\gb^2)^{-3\xi}\cdots$ have not 
been written out. We see that, for all $\xi<0$, we have a finite limit 
$T(0;g^2,\al)$ for $\gb^2\rightarrow0$. For the approach to $ T(0;g^2,\al)$,
the one-loop 
$(\gb^2)^{-\xi}$-term is relevant for $|\xi|<1$, otherwise the two-loop
 $\gb^2$-term involving $\ga_{F10}$ becomes important.

So far, we have obtained the asymptotic behavior of the function $T$ for 
$\gb^2\rightarrow+0$ provided $\xi<0$. With $T=-k^2B(k^2)$, we obtain then for 
$k^2\rightarrow-\infty$ along the negative real $k^2$-axis:
\be
-k^2B(k^2,\ka^2,g,\al)&\simeq& T(0;g^2,\al)\left\{1-\al\frac{\ga_{F01}}{\ga_{00}}
C_R^{-1}\left(-\ba_0\ln\frac{k^2}{\ka^2}\right)^{\xi} +\cdots\right.\nonumber\\
& & \left.+\frac{\ga_{F10}}{\ba_0}\left(-\ba_0\ln\frac{k^2}{\ka^2}\right)^{-1}+\cdots\right\} .
\label{19}
\en

In order to obtain the 
asymptotic properties for all directions in the complex $k^2-$plane, 
we use Eqs. (\ref{7q}) and (\ref{Tphi-}) and find that Eq.(\ref{19}) is also valid in all directions of the $k^2-$plane.

In view of the vanishing limit of $B(k^2)$ in (\ref{19}) for $k^2\ra\infty$ in all directions, we can write an unsubtracted 
dispersion representation for $B(k^2)$ :
\be
B(k^2) = \int^\infty_{-0}dk'^2\frac{\rho_B(k'^2)}{k'^2-k^2}.
\label{dispT}
\en
where $\rho(k^2)$ is the discontinuity of the function $B(k^2)$. From 
Eq.(\ref{19}), we also obtain the asymptotic expansion of the discontinuity of 
$B(k^2)$ for $k^2>0$ as 
\be
-k^2\rho_B(k^2,\ka^2,g,\al)&\simeq&T(0;g^2,\al)\left\{-\ga_{F01}
C_R^{-1}\al\left(-\ba_0\ln\frac{k^2}{|\ka^2|}\right)^{-1+\xi} +\cdots\right.\nonumber\\
& & \left.+\ga_{F10}\left(-\ba_0\ln\frac{k^2}{|\ka^2|}\right)^{-2}+\cdots\right\} .
\label{21}
\en
Since $-k^2B(k^2)\ra T(0;g^2,\al)$ in all directions 
in the complex $k^2-$plane, 
we obtain, from the unsubtracted dispersion relation (\ref{dispT}), the sum rule
\be
\int^\infty_{-0}dk^2\rho_B(k^2,\ka^2,g,\al) = T(0;g^2,\al) .
\label{22}
\en
It is important to emphasize, that equal-time commutation relations have {\it not} been
used in our arguments. Sum rules similar to Eq.({\ref{22}) are usually obtained
in field theory on the basis of these  relations, and with additional assumptions \cite{kl}. 
But equal time limits are very delicate in general field theories \cite{ku}.

We conclude that, for $\xi<0 ~~(10\leq N_F\leq16$ in QCD), the structure function 
$-k^2B(k^2)$ approaches a constant for all $\al\geq0$. Note that all our formulae 
are valid for $\al=0$, where the situation simplifies considerably, because there is
no dependence of the renormalization group equations upon the gauge field propagator.
In the Landau gauge $\al=0$, the asymptotic expressions (\ref{19}) and (\ref{21}) are 
independent of the parameter $\xi$. 

We complete the discussion of the case $\xi<0$ with the asymptotic expressions for 
the function $A(k^2)$ defined in Eq.(\ref{2}). For this purpose, we return to 
Eq.(\ref{8}).
For $k^2\rightarrow \infty$, we have $\ab\rightarrow0$ for the case $\xi<0$, 
and $\gb^2(|k^2/\ka^2|,g^2)\simeq(-\ba_0\ln\frac{k^2}{\ka^2})^{-1}\ra+0$. Since we have assumed 
that there are no mass parameters in the action, the function $S$ on the right 
hand side of Eq.(\ref{8}) should vanish for $\gb^2\rightarrow0$. The details of
this limit depend upon the specifics of possible mass generation, which we 
do not discuss here. We simply 
assume that $S(\frac{k^2}{\ka^2},g,\al)\simeq(g^2)^\lambda S_0(\al)$ for 
$g^2\rightarrow0$, where we consider $\lambda\geq1$. With this Ansatz,
we obtain from Eq.(\ref{8}), in the limit $k^2\rightarrow\infty$, the leading term 
\be
-k^2A(k^2,\ka^2,g,\al) \simeq T(0;g^2,\al)S_0(0)\left(-\ba_0 \ln
\frac{k^2}{\ka^2}\right)^{-\lambda}+\cdots ,
\label{24}
\en
and for the discontinuity of the function $A(k^2)$, we obtain, for $k^2>0$,
the leading term
\be
-k^2\rho_A(k^2,\ka^2,g,\al) \simeq T(0;g^2,\al)S_0(0)(-\ba_0\lambda)
\left(-\ba_0\ln\frac{k^2}{|\ka^2|}\right)^{-1-\lambda}+\cdots ,
\label{25}
\en
with $S_0(0)=S_0(\ab \rightarrow 0)$.
We again can write an unsubtracted dispersion relation for $A(k^2)$, 
\be
A(k^2) = \int^\infty_{-0}dk'^2\frac{\rho_A(k'^2)}{k'^2-k^2}.
\label{dispA}
\ee
Furthermore, since $k^2A(k^2)$
vanishes for $k^2\rightarrow\infty$, actually 
for any $\lambda>0$, we have the superconvergence relation
\be
\int^\infty_{-0}dk^2\rho_A(k^2,\ka^2,g,\al) = 0 .
\label{26}
\en
In the presence of mass generation, this relation  expresses the absence of 
a mass parameter in the original action \cite{ni}.

\bigskip
\bigskip

${\bf \gamma_{00}/\ba_0 > 0}$ : We now turn to the case $\xi\equiv\frac{\ga_{00}}{\ba_0}>0$,
 i.e. $\ga_{00}<0$, $ \ba_0<0$  ($0\leq N_F \leq 9$ in QCD). Here we have to 
require $\al>0$. As will be seen, the limit
$\al \rightarrow 0 $ of the asymptotic expressions does not exist in this case. 
Because of our requirement of asymptotic freedom, $\beta_0<0$, we generally 
have $\xi<1$ for QCD and similar theories. 

>From the expansion of $\ab$ given in Eq.(\ref{absoln}), we have the following expansion for $\frac{\ga_F(\gb^2,\ab)}{\beta(\gb^2)}$:
\be
\frac{\ga_F(\gb^2,\ab)}{\beta(\gb^2)}&\simeq& 
    \frac{\az\ga_{F01}}{\beta_0} (\gb^2)^{-1}
    + \frac{C\ga_{F01}}{\beta_0}(\gb^2)^{1-\xi} + O((\gb^2)^{1-2\xi})
    \nonumber\\
    && +\left(\frac{\ga_{F1}(\az)}{\beta_0}
        -\frac{\az\ga_{F01}\beta_1}{\beta_0^2}
        +\frac{\ga_{F01}\az\phi_0(\az)}{\beta_0(\xi-1)}\right)\nonumber\\
    && + \left(\frac{\gamma_{F01}C\Phi(\az)}{\beta_0} + 
        \frac{C\gamma_{F1}'(\az)}{\beta_0}-
        \frac{C\gamma_{F01}\beta_1}{\beta_0^2}\right)(\gb^2)^\xi
        \nonumber\\
    &&+ \cdots
\label{integrand}
\en
The first leading term indicates that $\frac{\ga_F(x,\ab(x))}{\beta(x)}$ is not integrable at $x=0$, but the second leading term shows that $\frac{\ga_F(x,\ab(x))}{\beta(x)}-
\frac{\az\ga_{F01}}{\beta_0x}$ is. We can therefore write Eq.(\ref{14}) as
\be
T(\gb^2;g^2,\al) &=&\exp\left(\int^{\gb^2}_{g^2}dx\frac{\ga_{F01}\al_0}{\beta_0x}\right) 
    \exp\left(\int^{\gb^2}_{g^2}dx\left(\frac{\ga_F(x,\ab(x))}{\ba(x)}-
    \frac{\ga_{F01}\al_0}{\beta_0x}\right)\right) \cr
    &=& \left(\frac{\gb^2}{g^2}\right)^{-\frac{\ga_{F01}}{\ga_{01}}\xi}\exp\int^0_{g^2}dx\left(\frac{\ga_F(x,\ab(x))}{\ba(x)}-
\frac{\ga_{F01}\al_0}{\beta_0x}\right) \nonumber\\
& &
\times\exp\int^{\gb^2}_{0}dx\left(\frac{\ga_F(x,\ab(x))}{\ba(x)}-
\frac{\ga_{F01}\al_0}{\beta_0x}\right). 
\en
Expanding the last integral for $\gb^2\ra0$ with Eq.(\ref{integrand}), we obtain for the asymptotic expansion of the quark structure function:
\be
T(\gb^2;g^2,\al) &\simeq& C_T(g^2,\al)\left\{(\gb^2)^{-\frac{\ga_{F01}}{\ga_{01}}\xi}
+ \frac{C\ga_{F01}}{\ga_{00}}(\gb^2)^{-\frac{\ga_{F01}}{\ga_{01}}\xi+
\xi}+\cdots \right. \nonumber\\
&&+\left.\left( \frac{\ga_{F1}(\al_0)}{\ba_0}-  
\frac{\ba_1}{\ba_0}\ga_{F01}\al_0 
+ \frac{\ga_{F01}\al_0 \phi_0(\al_0)}{\ba_0 (\xi-1)}\right)
({\gb^2})^{1-\frac{\ga_{F01}}{\ga_{01}}\xi}+\cdots\right\},\nonumber\\
\label{29}
\en
where the 
coefficient $C_T(g^2,\al)$ is defined as 
\be
C_T(g^2,\al)=(g^2)^{\frac{\ga_{F01}}{\ga_{01}}\xi}
\exp\int^0_{g^2}dx\left(\frac{\ga_F(x,\ab(x))}{\ba(x)}-
\frac{\ga_{F01}\al_0}{\ba_0x}\right) .
\label{30}
\en
With $\xi<1$ and $\ga_{F01}/\ga_{01}<1$ for the theories of interest
($\ga_{F01}/\ga_{01}=8/9$ in QCD), 
only the leading term is singular for $\gb^2\rightarrow0$ and hence for 
$k^2\rightarrow\infty$. 

With Eq.(\ref{29}), we  obtain in the case $\xi>0$ , $\al>0$, 
and in the limit $k^2 \rightarrow \infty$,
\be
-k^2B(k^2,\ka^2,g^2,\al) &\simeq& C_T(g^2,\al)\left(-\ba_0\ln\frac{k^2}{\ka^2}
\right)^{\frac{\ga_{F01}}{\ga_{01}}\xi}+\cdots,
\label{31}
\en
and correspondingly for the discontinuity
\be
-k^2
\rho_B(k^2,\ka^2,g^2,\al) &\simeq& C_T(g^2,\al)\gamma_{00}
\frac{\ga_{F01}}{\ga_{01}}\left(-\ba_0\ln\frac{k^2}{|\ka^2|}\right)^
{-1+\frac{\ga_{F01}}{\ga_{01}}\xi}+\cdots .
\label{32}
\en
Since $-k^2B(k^2)$ diverges  for $\xi>0$, we have no sum rule for $\al>0$. 
But since $B(k^2)\ra0$, the unsubtracted dispersion representation (\ref{dispT}) is certainly valid also 
for $\xi>0,\al>0$. 

Let us now consider $-k^2A(k^2,\ka^2,g,\al)=S(\frac{k^2}{\ka^2},g,\al)$ 
for $\xi>0,\al>0$. We again use Eq.(8), and find for the leading term,
\be
-k^2A(k^2) \simeq C_T(g^2,\al)S_0(\al_0)
\left(-\ba_0\ln\frac{k^2}{|\ka^2|}\right)^{-\lambda+\frac{\ga_{F01}}{\ga_{01}}\xi}+\cdots.
\label{33}
\en
We certainly have superconvergence if $\lambda\geq1$, as we have assumed, since 
$\xi\frac{\ga_{F01}}{\ga_{01}}<1$ in the theories considered. The discontinuity
$\rho_A(k^2)$ then satisfies again the superconvergence relation (\ref{26}), and it
has the asymptotic limit
\be
-k^2\rho_A(k^2) &\simeq &C_T(g^2,\al)S_0(\al_0)\ba_0
\left(-\lambda +\frac{\ga_{F01}}{\ga_{01}}\xi\right) 
\left( -\ba_0\ln\frac{k^2}
{|\ka^2|}\right)^{-\lambda-1+\frac{\ga_{F01}}{\ga_{01}}\xi} \cr
& &+ \cdots .
\label{34}
\en 

As we have pointed out, the asymptotic expression for $B(k^2)$ and $A(k^2)$ 
in the case $\xi<0$  ($10\leq N_F\leq16$ for QCD) are valid for 
the special case $\al=0$ 
(Landau gauge), while those for $\xi>0$  ($0\leq N_F\leq 9$) do not allow 
the limit $\al\rightarrow0$. In fact, for $\al=0$, the structure functions 
are independent of $\xi$. This parameter only enters the renormalization group 
equation via $\al$, which is not renormalization group invariant parameter. 

\bigskip
\bigskip

${\bf \ga_{00}=0}$ : It remains to consider the case $\xi=0$, i.e. $\ga_{00}=0$ 
and $\ba_0<0$,  for $\al\geq 0$. 
As has been pointed out, in the Landau gauge $\al=0$, we have no dependence of the 
quark propagator asymptotics upon $\xi$. Hence, for all 
values of $\xi$, the formulae (\ref{19}) for
$B(k^2)$ and (\ref{24}) for $A(k^2)$ are valid, setting $\al=0$, as are the sum
rules (\ref{22}) and (\ref{26}). 

For $\al>0$, the asymptotic expansion of the gauge field structure function is given in Eq.(\ref{D0}).
>From $\ab=\al R^{-1}$, we then have
\be
\ab &\simeq& -\frac{\beta_0}{\ga_{01}}\left(\ln\ln\frac{k^2}{\ka^2}\right)
^{-1} +\cdots\cr
    &\simeq& -\frac{\beta_0}{\ga_{01}}\left(\ln\gb^2\right)^{-1}+\cdots.
\label{ab0}
\ee
With this expression, the integrand in Eq.(\ref{14}) can be expanded as
\be
\frac{\ga_{F}(x,\ab(x))}{\beta_0{x}}\simeq
    \frac{\ga_{F01}}{\ga_{01}}\frac{1}{x\ln x}+\frac{\ga_{F1}(0)}{\beta_0}
    +\cdots.
\ee
The first two terms show that the function $\frac{\ga_{F}(x,\ab(x))}{\beta_0(x)}$ is non-integrable at $x=0$, but $\frac{\ga_{F}(x,\ab(x))}{\beta_0(x)} - \frac{\ga_{F01}}{\ga_{01}}\frac{1}{x\ln x}$ is. Hence we can write Eq.(\ref{14}) as 
\be
T(\gb^2;g^2,\al) &=&\exp\left(\int^{\gb^2}_{g^2}dx\frac{\ga_{F01}}{\ga_{01}}\frac{1}{x\ln x}\right)
    \exp\int^{\gb^2}_{g^2}dx\left(\frac{\ga_F(x,\ab(x))}{\ba(x)}-
    \frac{\ga_{F01}}{\ga_{01}}\frac{1}{x\ln x}\right) \nonumber\\
    &=& \left(\frac{\gb^2}{g^2}\right)^{-\frac{\ga_{F01}}{\ga_{01}}\xi}\exp\int^0_{g^2}dx\left(\frac{\ga_F(x,\ab(x))}{\ba(x)}-
\frac{\ga_{F01}\al_0}{\beta_0x}\right) \nonumber\\
& &
\times\exp\int^{\gb^2}_{0}dx\left(\frac{\ga_F(x,\ab(x))}{\ba(x)}-
\frac{\ga_{F01}\al_0}{\beta_0x}\right). 
\en
Expanding the last integral  for $\gb^2\ra0$, and keeping only the leading 
term, we find for $\xi=0$ and $\al>0$ in the limit $\gb^2\rightarrow0$:
\be
T(\gb^2;g^2,\al) &\simeq& C_0(g^2,\al)(\ln\gb^2)^{\frac{\ga_{F01}}{\ga_{01}}}+ \cdots ,
\label{40}
\en
and hence for $k^2\rightarrow\infty$:
\be
-k^2B(k^2)\simeq C_0(g^2,\al)\left(\ln\ln\frac{k^2}
{\ka^2}\right)^{\frac{\ga_{F01}}{\ga_{01}}}+\cdots ,
\label{41}
\en
The coefficient $C_0$ is given by
\be
C_0(g^2,\al) = (\ln g^2)^{-\frac{\ga_{F01}}{\ga_{01}}} \exp \int^0_{g^2} dx
\left( \frac{\ga_F(x,\bar{\al} (x))}{\ba(x)} - \frac{\ga_{F01}}
{\ga_{01}}\frac{1}{x\ln x}\right) .
\label{43}
\en
For the discontinuity $\rho_B$, we obtain the leading term, for real positive
$k^2\rightarrow+\infty$,
\be
-k^2\rho_B(k^2) &\simeq& -C_0(g^2,\al)\frac{\ga_{F01}}{\ga_{01}}\left(\ln\frac{k^2}
{|\ka^2|}\right)^{-1}
\left(\ln\ln\frac{k^2}{|\ka^2|}\right)^{\frac{\ga_{F01}}{\ga_{01}}-1}+\cdots .
\label{42}
\en
Since $-k^2B(k^2)$ diverges for $k^2\ra\infty$, there is no sum rule.

In order to obtain the asymptotic terms of the function $A(k^2)$ in the 
case $\xi=0$, $\al=0$, 
we can again Eq.(\ref{8}) and the Ansatz for the weak coupling limit
of the function $S$. We find
\be
-k^2A(k^2)\simeq C_0(g^2,\al)S_0(0)\left(-\ba_0\ln\frac{k^2}
{\ka^2}\right)^{-\lambda}
\left(\ln\ln\frac{k^2}{\ka^2}\right)^{\frac{\ga_{F01}}
{\ga_{01}}}+\cdots,~
\label{44}
\en 
and for the discontinuity,
\be
-k^2\rho_A(k^2)&\simeq &-C_0(g^2,\al)S_0(0)\ba_0\lambda\left
(-\ba_0\ln\frac{k^2}{|\ka^2|}\right)^{-\lambda-1}
\left(\ln\ln\frac{k^2}{|\ka^2|}\right)^{\frac{\ga_{F01}}
{\ga_{01}}}\cr
& &+\cdots.
\label{45}
\en 
We see that $A(k^2)$ again vanishes at infinity faster than $k^{-2}$, 
so that the superconvergence relation (\ref{26}) remains valid for  
$A(k^2)$ with $\xi=0, \al>0$.

\section{Ghost Propagator}

In this section, we present the asymptotic expansion of the ghost propagator for $k^2\ra\infty$. Our derivations will be brief since the essential steps are
identical to those used in the previous section for deriving the quark propagator.

The structure function of the ghost propagator is defined by 
\be
D_c(k^2+i0) = \int d^4x \exp^{ik\cdot x}
\langle 0|T\bar{\eta}(x)\eta(0)|0\rangle.
\label{D_c}
\en
We introduce the dimensionless function $R_c(\frac{k^2}{\ka^2},g,\al) = -k^2D_c(k^2,\ka^2,g,\al)$ and choose the normalization $R_c(1,g,\al)=1$.

With $\ka^2 < 0$ as the normalization point, we have the renormalization group equation for the ghost field
\be
\eta(x, g',\al',\ka'^2) = \sqrt{Z_c}\eta(x,g,\al,\ka^2)~,
\label{ghost}
\en
Here 
\be
Z_c= Z_c\left(\frac{\ka'^2}{\ka^2},g,\al\right),
\en
and the functions $g'$ and $\al'$ are defined in Eq.(\ref{4}).

By identical steps as in the previous two sections, we obtain the renormalization group equation for the ghost structure function
\be
R_c\left(\frac{k^2}{\ka^2},g,\al\right) &=& R_c\left(\frac{\ka'^2}
{\ka^2},g,\al\right)R_c\left(\frac{k^2}{\ka'^2},g',\al'\right) ,
\label{R_cnorm}
\en
and the corresponding differential equation
\be
\frac{\partial \ln R_c(\gb^2;g^2,\al)}{\partial\gb^2} = \frac{\ga^c(\gb^2,\bar\al)}{\ba(\gb^2)},
\label{R_cRGE}
\en
where the variable $\gb^2 = \gb^2(u,g^2)$ is defined as before, and $\ga^c(g^2,\al)\equiv u\frac{\partial R_c(u,g,\al)}{\partial u}|_{u=1}$ is the anomalous dimension of the ghost field. The asymptotic expansion of $\ga^c(g^2,\al)$ for $g^2\ra0$ is given by
\be
\gamma^c(g^2,\alpha) = \gamma^c_0 (\alpha) g^2 + \gamma^c_1 (\alpha) g^4
+\cdots,
\label{gammac}
\en 
where
\be
\gamma^c_0 (\alpha) &=& \gamma^c_{0 0} + \alpha \gamma^c_{01} ,\\
\gamma^c_1 (\alpha) &=& \gamma^c_{10} +
\alpha \gamma^c_{11} + \alpha^2\gamma^c_{12}.
\en
In QCD, 
\be
\gamma^c_{0 0} &=& -\frac{1}{16\pi^2}\frac{9}{4},\cr 
 \gamma^c_{01} &=& \frac{1}{16\pi^{2}}\frac {3}{4}.
\ee
As in the quark case, the solution to Eq.(\ref{R_cRGE}) can be written as
\be 
R_c(\gb^2;g^2,\al) = \exp\int^{\gb^2}_{g^2}dx\frac{\ga^c(x;\ab(x;g^2,\al))}{\ba(x)}.
\label{Rcint}
\en

To obtain the asymptotic expansion of $R_c(\gb^2)$ for $\gb^2\ra0$, we 
expand the integrand 
\be
\frac{\ga^c(\gb^2;\ab)}{\ba(\gb^2)}
\simeq \frac{\ga^c_0(\ab)}{\beta_0\gb^2} +
    \left(\frac{\ga_1^c(\ab)}{\beta_0}
        -\frac{\ga_0^c(\ab)\beta_1}{\beta_0^2}\right)
    +O(\gb^2)
\label{gabe}
\en
with the asymptotic expansions of $\ab(\gb^2)$
we have derived before for the cases of
$\xi<0$, $>0$ and $=0$. We discuss these three cases in the following separately.

\bigskip
\bigskip

${\bf \gamma_{00}/\beta_0 < 0} $ ($10\leq N_F\leq16$ in QCD):
Using the asymptotic expansion of $\ab(\gb^2)$ given in Eq.(\ref{absoln2}), we obtain
\be 
\f{\rc(\gb^2,\ab)}{\beta(\gb^2)}\simeq \f{\rc_{00}}{\beta_{0}\gb^2}
                +\f{C\rc_{01}}{\beta_{0}(\gb^2)^{1+\xi}}
        +\left(\frac{\ga_{10}^c}{\beta_0}-\frac{\ga_{00}^c\beta_1}{\beta_0^2}\right)+\cdots
\ee
The first term on the right hand side is the lowest order leading term. The
second lowest order term is given either by the second term or the third term depending on if $\xi<-1$.
These leading terms show that $\f{\rc(x,\ab(x))}{\beta(x)}$ is not integrable at $x=0$, but $\f{\rc(x,\ab(x))}{\beta(x)}-\f{\rc_{00}}{\beta_{0}x}$
is. We then obtain from Eq.(\ref{Rcint}) 
the asymptotic expansion for $R_c(\gb^2)$:
\be
R_c(\gb^2)\simeq C_{c-}(g^2,\al)(\gb)^{\f{\rc_{00}}{\beta_0}}\left\{1-\f{C\rc_{01}}{\ga_{00}}(\gb^2)^{-\xi}+\left(\frac{\ga_{10}^c}{\beta_0}-\frac{\ga_{00}^c\beta_1}{\beta_0^2}\right)+\cdots\right\},
\label{Rc-}
\en
where the coefficient $C_{c-}$ is given by
\be 
C_{c-}(g^2,\alpha) = (g^{2})^{-\f{\rc_{00}}{\beta_0}}
        \exp\int^{0}_{g^{2}}dx\left(
         \f{\rc(x,\ab(x))}{\beta(x)}-\f{\rc_{00}}{\beta_0x}\right).
\label{cg}
\ee
Eq.(\ref{Rc-}) then implies the following asymptotic expansion of the ghost structure function $-k^2D_c(k^2)$ for $k^2\ra-\infty$:
\be
& &-k^2D_c(k^2,\ka^2,g^2,\al) \simeq C_{c-}\left(-\ba_0\ln\frac{k^2}{\ka^2}\right)^{-\f{\rc_{00}}{\beta_0}} \nonumber\\
    & &~~~~\times\left\{1-\f{C\rc_{01}}{\ga_{00}}\left(-\ba_0\ln\frac{k^2}{\ka^2}\right)^{\xi}
    +\left(\frac{\ga_{10}^c}{\beta_0}-\frac{\ga_{00}^c\beta_1}{\beta_0^2}\right)\left(-\ba_0\ln\frac{k^2}{\ka^2}\right)^{-1}+\cdots\right\}.\nonumber\\
\label{Dc-}
\en

By the same steps we used in the previous sections, we can show that Eq.(\ref{Dc-}) is in fact valid for $k^2\ra\infty$ in all directions.
We then obtain for the expansion of the corresponding discontinuity:
\be
 -k^2\rho_c(k^2,\ka^2,g,\al) \simeq -C_{c-}\rc_{00}\left(-\ba_0\ln\frac{k^2}{|\ka^2|}\right)^{-\f{\rc_{00}}{\beta_0}-1}+ \cdots. 
\label{rhoc-}
\en

In QCD, $\rc_{00}/\beta_0=9/(\frac{11}{4}-\frac{1}{6}N_F)$. Since we require $\beta_0<0$ ($N_F\leq16$), $\rc_{00}/\beta_0>0$, so $-k^2D_c(k^2)\ra0$ as $k^2\ra\infty$ in all directions. We then have the unsubtracted dispersion relation for $D_c(k^2)$:
\be
D_c(k^2) = \int^\infty_{-0}dk'^2\frac{\rho_c(k'^2)}{k'^2-k^2}.
\label{dispRc}
\en
and the sum rule:
\be
\int_{-0}^{\infty}dk^2\rho_c(k^2,\ka^2,g^2,\al) =0.
\label{rhocsum}
\en

\bigskip
\bigskip

${\bf \gamma_{00}/\ba_0 > 0}$ ($0\leq N_F \leq 9$ in QCD):
If $\al=0$, then $\ab\equiv0$ and all the expressions in Eq.(\ref{Rc-}), (\ref{Dc-}), (\ref{rhoc-}), and (\ref{rhocsum}) are valid after setting $\al=0$.
If $\al>0$, then the asymptotic expansion of $\ab(\gb^2)$ in Eq.(\ref{absoln}) implies that 
\be \f{\rc(\gb^2,\ab)}{\beta(\gb^2)}\simeq 
            \f{\rc_{00}+\rc_{01}\az}{\beta_{0}\gb^2}
            +\f{C\rc_{01}}{\beta_{0}(\gb^2)^{1-\xi}}+\cdots
\ee
>From Eq.(\ref{Rcint}), we then obtain
\be
        R_{c} \simeq C_{c+}(g^2,\al)(\gb)^{\xi_{c}}\left\{1+\f{\rc_{01}C}{\beta_{0}\xi}(\gb)^{\xi}+\cdots\right\},
\en
where
\be
    \xi_{c}&=& \f{\rc_{00}+\rc_{01}\az}{\beta_{0}},\\
    C_{c+}(g^2,\alpha)&=& 
        (g^{2})^{-\xi_{c}}\exp\int^{0}_{g^{2}}dx\left(
         \f{\rc(x,\ab(x))}{\beta(x)}-\f{\xi_{c}}{x}\right) .
\ee
Then for the case $0<\xi<1$ the asymptotic expansion of $k^2D(k^2)$ for $k^2\ra\infty$ is given by
\be
-k^2D_c(k^2,\ka^2,g^2,\al)&\simeq &C_{c+}\left(-\ba_0\ln\frac{k^2}{\ka^2}\right)^{-\xi_c}\nonumber\\
    & &\times\left\{1+\f{C\rc_{01}}{\ga_{00}}\left(-\ba_0\ln\frac{k^2}{\ka^2}\right)^{-\xi}+\cdots\right\}.
\label{Dc+}
\en
The corresponding discontinuity at $k^2>0$ is given by 
\be
& &-k^2\rho_c(k^2,\ka^2,g,\al)\simeq -C_{c+}(\rc_{00}+\rc_{01}\az)
        \left(-\ba_0\ln\frac{k^2}{|\ka^2|}\right)
            ^{-\xi_c-1} \nonumber\\
    & &~~~~\times\left\{1-C\rc_{01}\left(\frac{1}{\ga_{00}}
            +\f{1}{\rc_{00}+\rc_{01}\az}\right)
    \left(-\ba_0\ln\frac{k^2}{|\ka|^2}\right)^{-\xi-1}+\cdots\right\}.\nonumber\\
\label{rhoc+}
\en

Eq.(\ref{Dc+}) shows that $D_c(k^2)$ vanishes at $k^2\ra\infty$ in all directions, hence the unsubtracted dispersion relation (\ref{dispRc}) still holds. Furthermore, in QCD, $\xi_c=(-1+\f{1}{3}N_{F})/(11-\f{2}{3}N_{F})$, so $\xi_c>0$ if $N_F>3$ ($N_F\leq9$ as required by $\xi>0$), $\xi=0$ if $N_F=3$, and $\xi<0$ if $N_F<3$. Eq.(\ref{Dc+}) then implies that as $k^2\ra\infty$, $k^2D(k^2)\ra0$ if $3<N_F\leq9$, $k^2D(k^2)\ra C_{c+}$ if $N_F=3$ and $k^2D(k^2)\ra\infty$ if $N_F<3$. Thus there are  the sum rules
\be
\int_{0+}^{\infty}dk^2\rho_c(k^2,\ka^2,g^2,\al) &=&0, ~~~~\mbox{for}~~4\leq N_F\leq9,\\
\int_{0+}^{\infty}dk^2\rho_c(k^2,\ka^2,g^2,\al) &=&C_{c+}(g^2,\al), ~~~~\mbox{for}~~N_F=3,
\label{rhocsum+}
\en
and there is no sum rule for $N_F<3$.

\bigskip
\bigskip

${\bf \ga_{00}=0}$ : 
Again, in the Landau gauge $\al=0$, we have no dependence of the 
quark propagator asymptotics upon $\xi$. Hence the  formula in Eq.(\ref{Rc-}), (\ref{Dc-}), (\ref{rhoc-}), and (\ref{rhocsum}) are still valid for $\xi=0$.

For $\al>0$, we obtain from the expansion (\ref{ab0}) of $\ab$ the expansion of $\ga^c/\beta$ as 
\be
\frac{\ga^c(\gb^2,\ab)}{\ba(\gb^2)}
\simeq \frac{\ga_{00}^c}{\beta_0\gb^2} +
    \frac{\ga_{01}^c}{\ga_{01}\gb^2\ln\gb^2}
        +\left(\frac{\ga_{10}^c}{\beta_0}
        -\frac{\ga_{00}^c\beta_1}{\beta_0^2}\right)
    +\cdots
\ee
This expansion shows that neither  $\frac{\ga^c(x,\ab(x))}{\ba(x)}$ 
nor $\frac{\ga^c(x,\ab(x))}{\ba(x)}- \frac{\ga^c_{00}}{\beta_0}$ 
is integrable at $x=0$, but the difference
$\frac{\ga^c(x,\ab(x))}{\ba(x)}-\frac{\ga_{00}^c}{\beta_0x} -
    \frac{\ga_{01}^c}{\ga_{01}x\ln x}$ is.
We then obtain from Eq.(\ref{Rcint}) 
\be
R_c(\gb^2) &\simeq& C_{c0}(g^2,\al)\left(\gb^2\right)^{\frac{\ga_{00}^c}{\beta_0}}
    \left(\ln\gb^2\right)^{\frac{\ga_{01}^c}{\ga_{01}}}\cr
    & &\times \left\{1-\frac{\ga_{01}^c}{\ga_{01}}\ln g^2\left(\ln\gb^2\right)^{-1}+\cdots\right\},
\ee
where 
\be
C_{c0}(g^2,\al) = \left(g^2\right)^{-\frac{\ga_{00}^c}{\beta_0}} 
    \exp\int_{g^2}^0dx \left(\frac{\ga^c(x,\ab(x))}{\ba(x)}-\frac{\ga_{00}^c}{\beta_0x} -
    \frac{\ga_{01}^c}{\ga_{01}x\ln x}\right).
\ee

The leading term of the ghost structure function $D_c(k^2)$ in the limit $k^2\ra\infty$ is then given by
\be
-k^2D_c(k^2) \simeq C_{c0}(g^2,\al)\left(-\ba_0\ln\frac{k^2}{\ka^2}\right)^{-\frac{\ga_{00}^c}{\beta_0}}
    \left(\ln\ln\frac{k^2}{\ka^2}\right)^{-\frac{\ga_{01}^c}{\ga_{01}}}
    +\cdots,
\ee
and the discontinuity
\be
-k^2\rho_c \simeq C_{c0}(g^2,\al)\ga_{00}^c\left(-\ba_0\ln\frac{k^2}{|\ka^2|}\right)^{-\frac{\ga_{00}^c}{\beta_0}-1}
    \left(\ln\ln\frac{k^2}{|\ka^2|}\right)^{-\frac{\ga_{01}^c}{\ga_{01}}}
    +\cdots.
\ee

In QCD, we have 
$\ga_{00}^c<0$ independent of the number of flavor $N_F$. Since
we require $\beta_0<0$, $\ga_{00}^c/\beta_0$ is positive. Thus $-k^2D_c(k^2)
\ra0$ in the limit $k^2\ra\infty$. We thus have the unsubtracted dispersion relation (\ref{dispRc}) and the sum rule
\be
\int_{-0}^{\infty}dk^2\rho_c(k^2,\ka^2,g^2,\al) =0.
\ee

\newpage

\section{Summary}

Finally we summarize the leading asymptotic terms for the structure 
functions of the  various propagators.
We use the following short-hand notation:
$\xi\equiv\frac{\ga_{00}}{\ba_0}$, 
$\xi_c\equiv\f{\rc_{00}+\rc_{01}\az}{\beta_{0}}$, $\al_0\equiv-\frac{\ga_{00}}{\ga_{01}}$,
$\ga_{F00}=0$;
$-k^2D(k^2,g^2,\ka^2.\al)=R(\frac{k^2}{\ka^2},g,\al)=
R(\gb^2;g^2,\al)=R(v)$, $\gb=\gb(v ,g^2) $, $v=\frac{k^2}{\ka^2}$, 
with corresponding relations for $B$ with $T$, $A$ with $S$ and $D_c$
with $R_c$ respectively.
We write $C_R=C_R(g^2,\al)$, and similarly for $C_T$, $C_0$ and other 
coefficients;   
$T(0)=T(0;g^2,\al)$, $R(0)=R(0;g^2,\al)$. $\rho=\rho(k^2,\ka^2,g,\al)$, 
$\rho_B$, $\rho_A$ and $\rho_c$ are the discontinuities of the corresponding structure
functions.

\ben
\mbox{\noindent {\bf $\xi<0:$}} &~& \ga_{00}<0, ~\ba_0<0, ~ (13N_C< 4N_F<22 N_C), ~\al\geq0 \\
& &  \\
& & R(v)\simeq C_R(-\ba_0\ln v)^{-\xi} +\cdots \\
& & T(v)\simeq T(0) + \cdots \\
& & S(v)\simeq T(0)S_0(0)(-\ba_{0}\ln v)^{-\lambda} + \cdots  \\
& & R_c(v)\simeq C_{c-}(-\beta_0\ln v)^{\f{\rc_{00}}{\beta_0}}+ \cdots  \\
& & \int^\infty_{-0}dk^2 \rho_B = T(0),~~~ \int^\infty_{-0}dk^2 \rho_A=0  ~~
(\lambda > 0)\\
& & \int^\infty_{-0}dk^2 \rho_c = 0
\enn

The coefficient is given by
\ben
& &C_{c-}(g^2,\alpha) = (g^{2})^{-\f{\rc_{00}}{\beta_0}}
        \exp\int^{0}_{g^{2}}dx\left(
         \f{\rc(x,\ab(x))}{\beta(x)}-\f{\rc_{00}}{\beta_0x}\right)
\enn

In the case $\al=0$, the results given above are valid for all $\xi \neq 0$.
 
\ben
\mbox{\noindent {\bf $\xi>0:$}} &~& \ga_{00}<0, ~\ba_0<0, ~ (0< 4N_F<13 N_C), ~ \al>0   \\
& &   \\
& & R(v)\simeq \frac{\al}{\al_0} + C_R(-\ba_0\ln v)^{-\xi} + \cdots \\
& & T(v)\simeq C_T(-\ba_{0}\ln v) ^{\frac{\ga_{F01}}{\ga_{01}}\xi} +\cdots\\
& & S(v)\simeq C_TS_0(\al_0)(-\ba_{0}\ln v)^{\frac{\ga_{F01}}{\ga_{01}}\xi-\lambda} 
 + \cdots\\
& & R_c(v)\simeq C_{c+}(-\beta_0\ln v)^{\xi_c} + \cdots\\
& & \int^\infty_{-0}dk^2 \rho = \frac{\al}{\al_0} , ~~~\int^\infty_{-0}dk^2 \rho_A=0 ~~(\lambda \geq 1) \\
& & \int^\infty_{-0}dk^2 \rho_c = 0 ~~(4\leq N_F\leq9)\\
& &  \int^\infty_{-0}dk^2 \rho_c = C_{c+} ~~(N_F=3) 
\enn

The coefficients are given by
\ben
& &C_T(g^2,\al)=(g^2)^{\frac{\ga_{F01}}{\ga_{01}}\xi}
\exp\int^0_{g^2}dx\left(\frac{\ga_F(x,\ab(x))}{\ba(x)}-
\frac{\ga_{F01}\al_0}{\ba_0x}\right)\\
& &C_{c+}(g^2,\alpha)= 
        (g^{2})^{-\xi_{c}}\exp\int^{0}_{g^{2}}dx\left(
         \f{\rc(x,\ab(x))}{\beta(x)}-\f{\xi_{c}}{x}\right)
\enn

\ben
\mbox{\noindent {\bf $\xi=0:$}}&~& \ga_{00}=0, ~\ba_0<0, ~~ (4N_F=13 N_C), ~~ \al>0    \\
& &     \\
& & R(v)\simeq -\al\frac{\ga_{01}}{\ba_0}\ln\ln v  +\cdots  \\
& & T(v)\simeq C_0(\ln\ln v) ^{\frac{\ga_{F01}}{\ga_{01}}} + \cdots \\
& & S(v)\simeq C_0(\ln\ln v)^{\frac{\ga_{F01}}{\ga_{01}}}(-\ba_{0}\ln v)^{-\lambda} + \cdots \cr 
& & R_c(v)\simeq C_{c0}(-\ba_0\ln v)^{-\frac{\ga_{00}^c}{\beta_0}}
    (\ln\ln v)^{-\frac{\ga_{01}^c}{\ga_{01}}}+ \cdots\\
& & \int^\infty_{-0}dk^2 \rho_A=0  ~~~(\lambda > 0)\\
& & \int^\infty_{-0}dk^2 \rho_c = 0
\enn

The coefficients are given by
\ben
& &C_0(g^2,\al) = (\ln g^2)^{-\frac{\ga_{F01}}{\ga{01}}} \exp \int^0_{g^2} dx
\left( \frac{\ga_F(x,\bar{\al} (x))}{\ba(x)} - \frac{\ga_{F01}}
{\ga_{01}}\frac{1}{x\ln x}\right) \\
& &C_{c0}(g^2,\al) = \left(g^2\right)^{-\frac{\ga_{00}^c}{\beta_0}} 
    \exp\int_{g^2}^0dx \left(\frac{\ga^c(x,\ab(x))}{\ba(x)}-\frac{\ga_{00}^c}{\beta_0x} -
    \frac{\ga_{01}^c}{\ga_{01}x\ln x}\right)
\enn

\ben
 \mbox{\noindent {\bf $\xi=0:$}}&~& \ga_{00}=0, ~\ba_0<0, ~~ (4N_F=13 N_C), ~~ \al=0   \\
& &     \\
& & R(v)\simeq R(0)\left(1 + \frac{\ga_{01}}{\ba_0}(-\ba_0\ln v)^{-1} + \cdots \right) \\
& & \int^\infty_0 dk^2 \rho = R(0)\\
\enn

With $\xi=0$, $\al=0$, the asymptotic terms for $T(v)$, $S(v)$ and $R_c(v)$ are obtained by setting $\al = 0$ 
in the case $\xi < 0$ given above.  

\bigskip
\bigskip

As can be seen from the asymptotic expressions summarized above, 
and the analytic properties discussed earlier, important aspects of the
structure functions are quite independent of the gauge parameter. 
Mainly as a consequence of asymptotic freedom, the functional form of the
asymptotic terms is determined by one-loop coefficients of the 
renormalization group equations. Generally, we find that the one-loop 
anomalous dimension coefficient $\gamma_{00}$ of the gauge field in the 
Landau gauge $\alpha=0$, corresponding to a fixed point of the 
function $\ab(u,g^2,\al)$ for $u\ra\infty$, plays an important r\^{o}le 
for {\em all} covariant gauges, and also for most structure functions.

In some cases, the sign of the coefficient $\ga_{00}$ decides the existence or
non-existence of sum rules for the discontinuities of structure functions.
More directly than the un-subtracted dispersion representations, these sum rules 
imply a connection between long- and short-distance aspects of the theory.

It is of interest to generalize our results to supersymmetric models in 
the Wess-Zumino gauge. Within this framework, one may find interesting
correlations between convergence properties and the phase structure of 
the SUSY models, which can be obtained from duality transformations. 
As has been mentioned in the introduction, some such connection has already
been explored \cite{on}.

\section*{Acknowledgments}

I am grateful to Professor R. Oehme for his guidance, inspiration and patience.

\end{document}